\documentclass[preprint]{aastex}
\usepackage{graphicx}

\def\etal{{\it et al.}~}

\begin{document}

\title{Planetesimal disk evolution driven by 
embryo-planetesimal  gravitational scattering.}

\author{R. R. Rafikov}
\affil{Princeton University Observatory, Princeton, NJ 08544}
\email{rrr@astro.princeton.edu}

\begin{abstract}
The process of gravitational scattering of planetesimals by a 
massive protoplanetary embryo is explored theoretically. 
We propose a method to
describe the evolution of the disk surface density, eccentricity, 
and inclination caused by the embryo-planetesimal interaction. It relies
on the analytical treatment of the scattering in two
extreme regimes of the planetesimal epicyclic velocities: shear-dominated 
(dynamically ``cold'') and dispersion-dominated (dynamically ``hot''). 
In the former, planetesimal scattering can be treated as a deterministic 
process. In the latter, scattering is mostly weak because of 
the large relative 
velocities of interacting bodies. This allows one to use the
Fokker-Planck approximation and the two-body approximation 
to explore the disk evolution. We compare the results obtained by 
this  
method with the outcomes of the direct numerical integrations of 
planetesimal
orbits and they agree quite well. 
In the intermediate velocity regime an approximate treatment of the
disk evolution is proposed based on interpolation between the two
extreme regimes.
We also calculate the rate of embryo's mass growth in 
an inhomogeneous planetesimal disk 
and demonstrate that 
it is in agreement with both the simulations and earlier calculations. 
Finally we discuss the question 
of the direction of the 
embryo-planetesimal interaction in the dispersion-dominated regime 
and demonstrate that it is repulsive.
This means that the embryo always forms a gap in the disk around
it, which is in contrast with the results of other authors.  The 
machinery developed here will be applied to realistic 
protoplanetary systems
in future papers.
\end{abstract}


%*************************************************************
%*************************************************************

\section{Introduction.}  
\label{sect:intro}

%*************************************************************
%*************************************************************

This paper continues the line of investigation started in 
our previous work (Rafikov 2001, 2002a; hereafter Papers I and II) which was 
devoted to the treatment of planetesimal-planetesimal 
gravitational interactions. Here we consider the interaction between 
the growing protoplanetary embryo and the planetesimal disk. By embryo we 
imply in the present context a single body
with mass $M_e$ much larger 
than the masses of individual planetesimals $m$.
There are several reasons for studying this important
problem separately from the mutual gravitational scattering of 
planetesimals.

First,
planetesimal-planetesimal encounters in realistic protoplanetary disks
usually occur in the dispersion-dominated regime, which applies
when the relative approach velocity of two particles is bigger than the 
differential shear in the disk across the Hill (or tidal) radius.
The Hill radius is defined as 
\begin{eqnarray}
r_H=a\left(\frac{m_1+m_2}{M_c}\right)^{1/3},
\label{eq:Hill_radius}
\end{eqnarray}
where $a$ is a value of the semimajor axis at which the 
interaction takes place, and $m_1,m_2,$ and $M_c$ 
are the masses of interacting
planetesimals and of the central star.

However, in the same protoplanetary disk gravitational interaction
between the embryo and planetesimals could be in the opposite 
velocity regime --- shear-dominated --- when the planetesimal 
random motion is small compared to the shear across the Hill radius,
simply because the embryo mass and therefore the Hill radius is much 
larger. Indeed, in the case of embryo-planetesimal interactions 
Hill radius $R_H=a_e(M_e/M_c)^{1/3}\gg r_H$ (here $a_e$ is a semimajor axis
of the embryo) as a consequence of 
$M_e\gg m_i$. Thus, reduced (normalized in Hill coordinates)
values of random velocities in the embryo-planetesimal case 
are smaller by a factor
$[(m_1+m_2)/M_e]^{1/3}\ll 1$ than those corresponding to the 
planetesimal-planetesimal interactions.

Of course, it might be that the 
planetesimal disk has  already been so  excited
dynamically that even embryo-planetesimal encounters are in the 
dispersion-dominated regime. Thus, we 
consider here both regimes of the embryo-planetesimal interaction
 --- shear- and dispersion-dominated. 

Second, embryo-planetesimal interactions are complicated by the 
presence of a special type of orbits in a $3$-body problem --- the
so-called
horseshoe (or librating) orbits 
(H\'enon \& Petit 1986; Murray \& Dermott 1999). 
Planetesimals on these orbits do not perform the usual 
 circulating motion which is
characteristic of passing orbits (the most important case for 
planetesimal-planetesimal scattering) but a librating one.
This horseshoe motion can only occur when the difference in 
semimajor axes of interacting bodies is smaller than their Hill radius. 
For planetesimal-planetesimal interactions
$r_H$ is negligible compared to the scale of surface density variations 
or the radial epicyclic excursion of an individual planetesimal. Thus 
horseshoe orbits are unimportant in this case. However, the Hill radius of
the embryo-planetesimal interaction $R_H$ can be comparable to the 
length scale of the disk inhomogeneities caused by the embryo.
Thus the phenomenon of horseshoe motion can be crucial for the
planetesimal dynamics near the embryo 
(see below \S \ref{subsubsect:hs_separation}). 

Third, as we have already mentioned in 
Paper II, planetesimal-planetesimal scattering is described in terms of 
the disk properties {\it averaged} over some region of the disk, which 
diminishes the importance of the details of spatial distributions of disk 
properties. On the contrary,
in the case of the embryo-planetesimal
interaction we are interested in details of the {\it spatial behavior} 
of various quantities characterizing the state of the disk and they are 
the primary goal of our present study.

All these complications preclude the direct application of 
the results obtained in
Paper II to the present consideration. However the general analytical 
approach to the treatment of planetesimal disk evolution 
developed there remains valid and we will employ it in this paper.

Numerical orbit integrations (Tanaka \& Ida 1996, 1997) and N-body 
simulations (Ida \& Makino 1993) 
provide an alternative and important route of studying 
embryo-planetesimal interactions. Their drawback is their intrinsically
low speed and inability to treat large number of planetesimals.
However, since the physics is incorporated in them on a very basic level 
with the minimum of additional assumptions  
they can provide us with robust predictions. To use this advantage 
of numerical methods and to avoid their handicaps we employ 
the following methodology: we provide a self-consistent 
{\it analytical} description of the 
embryo-planetesimal interaction in different velocity regimes. 
To check this description and to verify the validity of the 
simplifying assumptions utilized in its
development we have used numerical orbit integrations performed for 
several 
sets of typical planetesimal disk parameters. After we make sure that 
our theory works well for these sets of parameters we can use it for
others as well and be confident of its reliability. What we gain by this
approach is the speed of computation and ability to explore
the whole space of important physical variables.

The condition on the embryo's mass, $M_e\gg m$,
 has important dynamical  implications. In many applications it justifies the 
neglect of the embryo's recoil resulting from planetesimal  scattering.
Also, dynamical friction between the embryo and planetesimals will tend to 
produce random energy equipartition (Stewart \& Wetherill 1988; 
Wetherill \& Stewart 1989) which 
means that embryo's eccentricity and inclination are most likely to
be negligibly small. Thus, we will assume in this paper that the embryo moves 
on a fixed circular orbit and its eccentricity and inclination are zero.
We will also consider the embryo to be isolated from the gravitational 
effects of other massive bodies which may be growing nearby, an 
assumption which can easily be abandoned in future work.
Throughout the paper we neglect the presence of any resonant effects. This
is justifiable if frequent planetesimal-planetesimal encounters can 
destroy any commensurabilities with the embryo's rotation 
period. The embryo's recoil 
in the course of planetesimal scattering and distant 
embryo-embryo interactions 
would also help to do that. We leave for the future the clarification of
conditions necessary for employing this simplification.

It will be convenient to use the embryo's Hill radius $R_H$ as a unit of 
length in our study. We introduce the
Hill orbital elements of a planetesimal evaluated 
at large azimuthal distance from the embryo --- 
the difference in semimajor axes $H$, eccentricity 
$\tilde e$, and inclination $\tilde i$ relative to the embryo's orbit
at semimajor axis $a_e$:
\begin{eqnarray}
H=\frac{a-a_e}{R_H},~~~\tilde e=e\frac{a_e}{R_H},~~~\tilde i
=i\frac{a_e}{R_H},
\label{eq:new_vars}
\end{eqnarray}
and use them as planetesimal coordinates.
The distribution function of the orbital elements $\tilde e$ and $\tilde i$ 
will be assumed to have a Rayleigh form
 with dispersions $\tilde \sigma_e$ and $\tilde \sigma_i$:
\begin{equation}
\psi(\tilde e,\tilde i)d\tilde e d\tilde i=\frac{\tilde e d\tilde e 
~\tilde i d\tilde i}{\tilde \sigma_{e}^2
\tilde \sigma_{i}^2}\exp\left[-\frac{\tilde e^2}{2\tilde \sigma_{e}^2}
-\frac{\tilde i^2}{2\tilde \sigma_{i}^2}\right].
\label{eq:Railey}
\end{equation} 
We will also
be using the dimensionless surface number density of  
guiding centers $N(H)$ to characterize the 
planetesimal spatial distribution.

Although we focus on a single embryo, 
for many applications we can treat the embryo using a 
 continuous form of the evolution equations. Fro example, we may
assume that 
the discrete surface number density of the embryo is given by
\begin{eqnarray}
N_{em}(H)=\frac{1}{2\pi}\delta(m-M_e)\delta(H).
\label{eq:em_surf_density}
\end{eqnarray}
Since planetesimal-planetesimal interactions are not important here it will
be enough to consider a single-mass planetesimal population.

The three-body interaction in the Hill approximation preserves a certain 
combination of relative orbital elements of interacting bodies
called the Jacobi constant (Goldreich \& Tremaine 1980; H\'enon \& Petit 1986):
\begin{eqnarray}
J=\tilde e^2+\tilde i^2-\frac{3}{4}H^2+2\phi_e,
\label{eq:Jacobi}
\end{eqnarray}
where $\phi_e$ is the gravitational potential of the embryo, which can 
be neglected far from the embryo.
For embryo-planetesimal scattering one can introduce the
concept of integrated Jacobi constant of the whole 
planetesimal population:
\begin{eqnarray}
J^{tot}=\int\limits_{-\infty}^{\infty}
\left[2N(H)\tilde \sigma_{e}^2+2N(H)\tilde \sigma_{i}^2
-\frac{3}{4}N(H)H^2\right]dH.
\label{eq:int_Jacobi}
\end{eqnarray}
This quantity should be conserved  because
(1) each individual planetesimal scattering off the embryo conserves
the Jacobi constant of the relative motion, and (2) 
embryo's random motion is negligible, which means
that relative eccentricity, inclination, and difference in semimajor axes 
are determined by planetesimal orbital parameters only.
We will use the conservation of this quantity 
and of the total number of planetesimals  (we neglect their
coagulation at this point)
\begin{eqnarray}
N^{tot}=\int\limits_{-\infty}^{\infty}N(H)dH.
\label{eq:int_number}
\end{eqnarray}
as checks of our evolution equations.

The orbit integrations that we use are performed by solving
 Hill equations numerically.
We have integrated the evolution of the system [equations of 
osculating orbital elements evolution (11) of Paper II] using
fourth order Runge-Kutta integrator (Press \etal 1988). Unlike
similar calculations of  
Tanaka \& Ida (1996, 1997) our orbit integrations  do not employ  
additional analytical simplifications to avoid possible biases. 
In a typical integration the Jacobi constant is conserved with fractional 
accuracy $10^{-8}-10^{-12}$. The results of these orbit integrations and their 
comparisons with theoretical predictions will be
presented in the following sections.

We devote \S \ref{subsect:em_shear}
to studying the shear-dominated case and \S \ref{subsect:em_dispersion} to
exploring the dispersion-dominated case. 
The velocity regime intermediate between them 
is addressed in Appendix \ref{app:intermediate_velocity}.
We discuss some general features of the embryo-planetesimal interaction
in \S \ref{sect:repulsion}. Some auxiliary results 
are presented in appendices: Appendix \ref{app:scat_erobab} contains 
the derivation of the probability distribution 
of scattered semimajor axes in the 
dispersion-dominated regime, while in Appendix \ref{app:accr_rate} 
we calculate the embryo's accretion rate in different velocity regimes.

%*************************************************************
%*************************************************************

\section{Scattering by the embryo in the shear-dominated regime.}
\label{subsect:em_shear}

%*************************************************************
%*************************************************************

We consider first the embryo-planetesimal interaction in the
shear-dominated regime. In Paper II we have derived a master
equation (30) for the evolution of the planetesimal distribution 
function. We will use this general equation and equation (33) of Paper II
as a basis for our further developments.
The conditions for the shear-dominated regime to be realized are
\begin{eqnarray}
\tilde \sigma_{e}^2\ll 1,~~~~~~ 
\tilde \sigma_{i}^2\ll 1.
\end{eqnarray}
%where subscript ``j'' refers to planetesimals with mass $m_j$.
Some important simplifications can then be made.

First, scattering in this regime is  deterministic in the
sense that the outcome of an interaction between two particles 
depends only on their difference in semimajor axes before 
the collision
$H_0$ and not on their relative random motion (which is naturally 
absent in this case). This means that the change of the reduced 
semimajor axis difference $\Delta \tilde h_{sc}$\footnote{We use this 
notation instead of $\Delta H_{sc}$ 
to parallel the discussion of Paper II.} is a 
single-valued function of only $H_0$ in this regime. 

Second, inclination is hardly excited at all in the course of an encounter.
A considerable change of inclination requires a substantial 
force acting in the vertical direction. However, the dynamically  cold
disk is very thin and it is easy to see 
that the gravitational force between the interacting particles is directed
almost horizontally. Thus, noticeable inclination growth can occur only
for particles experiencing very close encounters. 
From qualitative considerations
one would expect that change of the inclination vector  ${\bf\tilde i}$ 
(see Petit \& H\'enon 1986; Ida 1990; Paper II)
is given by
\begin{eqnarray}
\Delta ({\bf\tilde i}_{sc})^2=
{\bf\tilde i}_0^2 ~g_1(H_0)\ll 1,
\label{eq:di_cold}
\end{eqnarray}
where ${\bf\tilde i}_0$ is the initial value of the 
reduced vector inclination and 
$g_1(H_0)\sim 1$ is some function which can be easily computed 
numerically. For our purposes we
neglect the inclination growth due to the gravitational stirring
in the shear-dominated regime completely
and set $\Delta {\bf\tilde i}_{sc}=0$. We will however keep the 
terms describing the transport of vertical energy in the disk (see below).

The absence of heating in the vertical direction 
naturally leads to another simplification.
Since we can neglect the change of inclination in the encounter
the change of eccentricity becomes directly related to the
change of semimajor axis difference due to the conservation
of Jacobi constant (\ref{eq:Jacobi}).
Then we can write that the change of the vector eccentricity  
${\bf\tilde e}$ is
\begin{eqnarray}
\Delta ({\bf\tilde e}_{sc})^2=
\frac{3}{4}\Delta (\tilde h_{sc}^2)=\frac{3}{4}\left[
(\Delta \tilde h_{sc})^2+2H_0
\Delta \tilde h_{sc}\right]. 
\label{eq:de_cold}
\end{eqnarray}
Note that $\Delta ({\bf\tilde e}_{sc})^2\sim 1$ if $H_0\sim 1$.
It can also be easily shown that 
\begin{eqnarray}
{\bf\tilde e}\cdot\Delta {\bf\tilde e}_{sc}\sim {\bf\tilde e}^2\ll 1,~~~~
{\bf\tilde i}\cdot\Delta {\bf\tilde i}_{sc}\sim {\bf\tilde i}^2\ll 1
\label{eq:friction}
\end{eqnarray}
in the shear-dominated regime; the corresponding scattering coefficients in 
evolution equations will therefore be neglected in this paper.

Thus, the probability $\tilde P_r$ of scattering 
from $H_0,{\bf\tilde e}_0,{\bf\tilde i}_0$ 
to $H,{\bf\tilde e},{\bf\tilde i}$
can be written in the shear-dominated regime as
\begin{eqnarray}
\tilde P_r=
\delta[\Delta H-\Delta\tilde h_{sc}
(H_0)]~\delta[\Delta{\bf \tilde e}-\Delta{\bf \tilde e}_{sc}
(H_0)]
~\delta(\Delta{\bf \tilde i}),
\label{eq:delta_fun_cold}
\end{eqnarray}
where $\delta$ denotes the Dirac delta function.

The computational challenge 
 of the shear-dominated regime is that strong scattering 
is possible for $H_0\sim 1$, 
i.e. $\Delta H$ in the course of an encounter 
can be quite substantial. Thus we cannot use the Fokker-Planck 
formalism for the shear-dominated scattering
and one has to deal with the different moments of the 
master evolution equation 
(30) of Paper II in their general form.
However, this does not
pose an insurmountable problem because the probability distribution function
of the shear-dominated scattering is a single-variable function only. 
An analytical fit to this function 
was calculated by Petit \& H\'enon (1986, 1987b) 
using results of numerical orbit integrations. 
This fit automatically
takes horseshoe motion into account so that we do not have to worry
about the complications associated with this type of orbit 
 in the shear-dominated regime: 
one can see from the expression
for $\Delta \tilde h_{sc}(H_0)$ (Petit \& H\'enon 1987b)
that $\Delta \tilde h_{sc}(H_0)\to -2H_0$ in the
shear-dominated regime as $H_0\to 0$.

Using the deterministic 
form of $\tilde P_r$ [substituting expression (\ref{eq:delta_fun_cold}) into
equation (30) of Paper II] and the embryo's surface density in the form
(\ref{eq:em_surf_density})
 we can take different moments of $\tilde {\bf e}^2$ and 
$\tilde {\bf i}^2$. As a result we find the following set of 
equations describing the evolution of the planetesimal population 
due to the scattering by the embryo in the shear-dominated regime:
\begin{eqnarray} 
&& \frac{\partial N(H)}{\partial t}=
-\frac{|A|\mu_e^{1/3}}{\pi}\left[N(H)|H|-
\int\limits_{-\infty}^{\infty}
\tilde P(H_0,H)N(H_0)|H_0|dH_0\right],
\label{eq:surf_shear}\\
&& \frac{\partial }{\partial t}\left[2N(H)\tilde\sigma_{e}^2(H)\right]=
-\frac{|A|\mu_e^{1/3}}{\pi}\Bigg[2N(H)\tilde\sigma_{e}^2(H)|H|
\nonumber\\
&& -\int\limits_{-\infty}^{\infty}
\tilde P(H_0,H)N(H_0)\left(2\tilde\sigma_{e}^2(H_0)+
\Delta ({\bf \tilde e}_{sc})^2(H_0)\right)|H_0|dH_0\Bigg],
\label{eq:ecc_shear}\\
&& \frac{\partial}{\partial t}\left[2N(H)\tilde\sigma_{i}^2(H)\right]=
-\frac{|A|\mu_e^{1/3}}{\pi}\nonumber\\
&& \times\left[2N(H)\tilde\sigma_{i}^2(H)|H|-
\int\limits_{-\infty}^{\infty}
\tilde P(H_0,H)2N(H_0)\tilde\sigma_{i}^2(H_0)
|H_0|dH_0\right],
\label{eq:inc_shear}
\end{eqnarray}
with $\mu_e=(M_e/M_c)^{1/3}$.
Here $A$ is the Oort's constant characterizing the differential rotation 
of the disk (Binney \& Tremaine 1987), 
$A=-(3/4)\Omega$ in a Keplerian disk ($\Omega=\sqrt{GM_c/a_e^3}$
is the disk rotation frequency),
$H_0$ is an integration variable having the meaning of the 
initial difference of the semimajor axes of interacting bodies, 
$\Delta ({\bf \tilde e}_{sc})^2$ is defined in 
equation (\ref{eq:de_cold}), and 
\begin{eqnarray}
\tilde P(H_0,H)=\delta\left[H-H_0-\Delta \tilde h_{sc}(H_0)\right].
\end{eqnarray}

Terms in the r.h.s. of equation (\ref{eq:inc_shear}) are 
$\sim \tilde \sigma_{i}^2\ll 1$, i.e. are of the same order as 
$\Delta ({\bf\tilde i}_{sc})^2$ which we have agreed to neglect. This
inconsistency stems from the fact that we want our evolution 
equations to preserve the integrated Jacobi constant (\ref{eq:int_Jacobi}). 
For this reason we keep
the terms in the r.h.s. of (\ref{eq:inc_shear}) (which are essentially 
the transport terms) but neglect terms like 
$\Delta ({\bf\tilde i}_{sc})^2$
if $\Delta ({\bf\tilde e}_{sc})^2$ is assumed to be given by
(\ref{eq:de_cold}). Then $\partial J^{tot}/\partial t=0$ which
can be verified using equations (\ref{eq:de_cold}) and
(\ref{eq:surf_shear})-(\ref{eq:inc_shear}). Also, one can easily
check that equation (\ref{eq:surf_shear}) conserves the total number
of planetesimals in the disk.

One technical issue merits mentioning at this point. 
In the shear-dominated regime the semimajor axis difference after 
the encounter $H$ is a single-valued function of $H_0$.  However,
the inverse function $H_0(H)$ is multivalued (Petit \& H\'enon 1987b). 
For this reason
integrals of some quantity $F(H_0)$ [e.g. $F=N(H_0)|H_0|$]
over $\tilde P(H_0,H)dH_0$ in (\ref{eq:surf_shear})-(\ref{eq:inc_shear})
result in
\begin{eqnarray}
\int\limits_{-\infty}^{\infty}
\tilde P(H_0,H)F(H_0)dH_0=\sum\limits_k 
F(H_{0 k})\left[1+\frac{\partial \Delta \tilde h_{sc}}{\partial H_0}
\right]^{-1},
\label{eq:integration_rule}
\end{eqnarray}
where $H_{0 k}$ is the $k$-th root of 
the equation\footnote{This form 
arises because of the integration rule of the Dirac 
$\delta$-function:
$\int\limits_{-\infty}^\infty p(x)\delta[q(x)]=
\sum\limits_{j}p(x_j)/q^\prime(x_j)$, where $x_j$ is the $j$-th root
of equation $q(x)=0$.}
\begin{eqnarray}
H=H_0+\Delta \tilde h_{sc}(H_0).
\label{eq:roots}
\end{eqnarray}

The system (\ref{eq:surf_shear})-(\ref{eq:inc_shear}) forms a closed set of 
equations needed to describe the disk evolution 
caused by embryo-planetesimal
scattering in the shear-dominated regime. Equation  (\ref{eq:surf_shear})
has already been derived by a different method in Paper I, 
and now we have extended that analysis
by taking random velocity evolution of the disk 
into account. Following Paper I
we will be using $\Delta \tilde h_{sc}(H_0)$ in the analytical form
suggested by Petit \& H\'enon (1987b; see also Appendix B of Paper I).

%*************************************************************
%*************************************************************

\section{Scattering by the embryo in the dispersion-dominated regime}
\label{subsect:em_dispersion}

%*************************************************************
%*************************************************************

When the embryo-planetesimal scattering is in the dispersion-dominated regime
the same simplifications as in the case of planetesimal-planetesimal 
scattering can be made: scattering is weak and 
this warrants the use of both the two-body
approximation and the  Fokker-Planck expansion. 
In this regime we can use many of the results 
obtained in Paper II. 
There are however some additional complications; one of them is related to the 
aforementioned existence of horseshoe orbits.
To discuss this issue we need to know which conditions 
should be fulfilled for 
horseshoe motion to take place in the dispersion-dominated 
velocity regime.

%%%%%%%%%%%%%%%%%%%%%%%%%%%%%%

\subsection{Horseshoe orbits.}
\label{subsubsect:hs_separation}

%%%%%%%%%%%%%%%%%%%%%%%%%%%%%%

\begin{figure}[t]
\vspace{12.cm}
\includegraphics{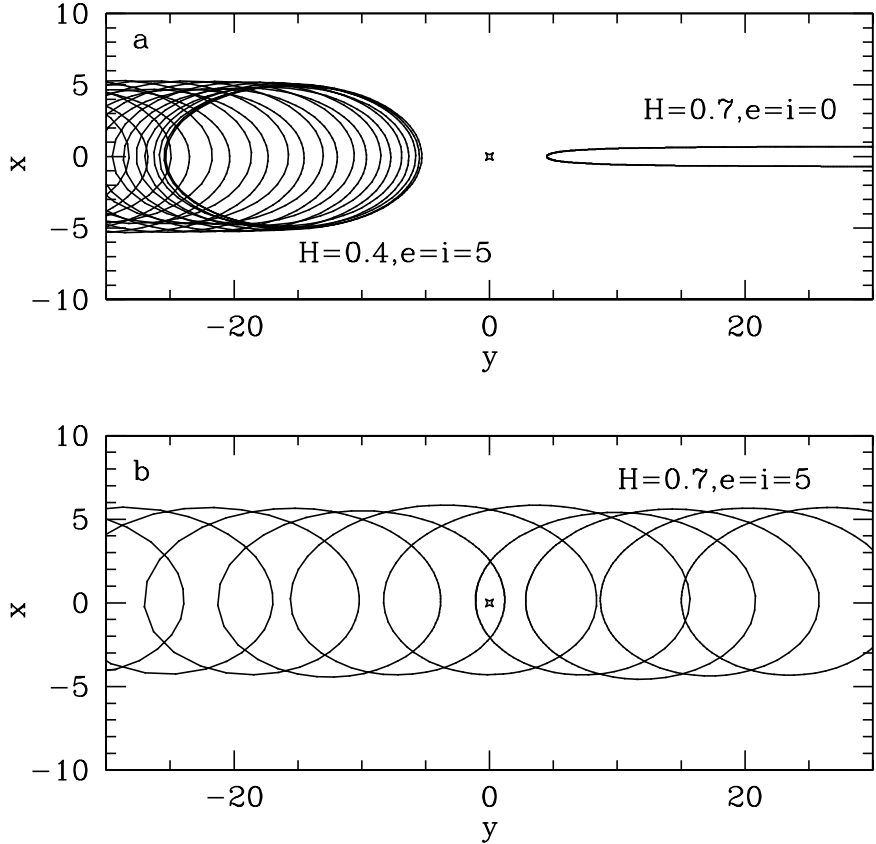}
\caption{Representative trajectories of bodies on horseshoe ({\it a}) and
passing ({\it b}) orbits for different values of initial orbital elements.
The scattering body is at the origin and the coordinates are in units 
of the Hill radius.}
\label{fig:plot_orbits}
\end{figure}

Horseshoe motion arises when the separation of the 
semimajor axes of interacting 
bodies is small. In this case their relative motion
due to the shear in the disk is very slow; for this reason 
even the weak 
gravitational force between the bodies at large azimuthal 
distance can lead
to a considerable change of their angular momenta over a long time interval;
the relative motion of particles reverses, they
turn around and move on almost closed trajectories until the next 
conjunction (see Figure \ref{fig:plot_orbits}a).
In the Solar System the best example of such motion is given by Saturnian 
satellites Janus \& Epimetheus (Murray \& Dermott 1999).

In contrast, planetesimals on 
pasing orbits pass each other after the interaction, usually
without reversing their motion
(see Figure \ref{fig:plot_orbits}b). In the shear-dominated regime 
these orbits
typically have initial separations of semimajor axes larger than
the Hill radius. However in dispersion-dominated encounters 
the question of the spatial separation of horseshoe and passing orbits 
becomes more subtle.

Namouni (1999) has considered dispersion-dominated
scattering in the planar case ($\tilde i=0$) and
suggested that for a given relative eccentricity $\tilde e\gg 1$ 
the boundary of the horseshoe region is located at
$H=\tilde h_{hs}\sim \tilde e^{-1/2}$ 
[see also Ida \& Nakazawa (1989)]. This means that the more ``energetic''
the planetesimals are, the smaller their semimajor axis separation has to be 
for horseshoe motion to appear. 
This tendency is illustrated in Figure \ref{fig:plot_orbits}: orbits with the
same value of $H=0.7$ exhibit different behavior depending on their
random velocities --- in the shear-dominated regime ($\tilde e=\tilde i=0$)
horseshoe motion takes place while in the dispersion-dominated 
($\tilde e=\tilde i=5$) the orbit is passing; however if one decreases $H$ 
this orbit also becomes horseshoe (which is
shown by a trajectory with $H=0.3,\tilde e=\tilde i=5$ in Figure 
\ref{fig:plot_orbits}a).
It is easy to understand the
 reason for this dependence on the planetesimal  random velocity:
the higher the eccentricity the more time the 
approaching planetesimal spends
far from the scatterer because it moves 
on a very large epicycle. As a result the 
mutual force  is weak, the angular momentum  
exchange is small, and the planetesimal passes the 
scatterer instead of turning around.
In other words, the higher $\tilde e$ the less noticeable the presence 
of the scatterer is for the incoming planetesimal.

\begin{figure}[t]
\vspace{11.cm}
\includegraphics{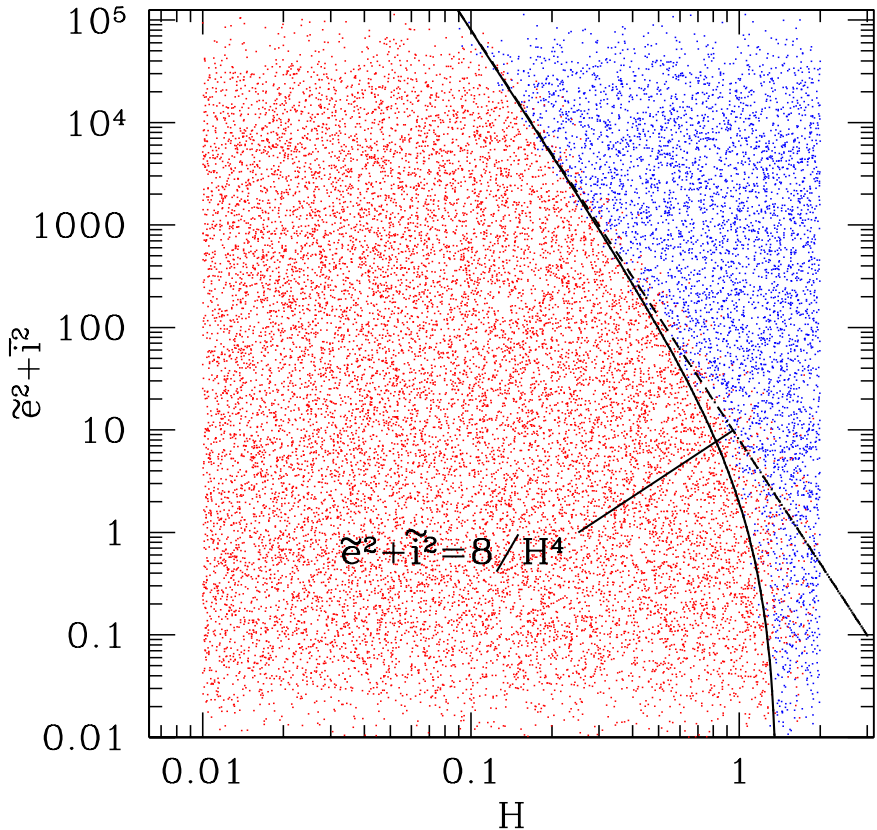}
\caption{Horseshoe and passing orbits as a function of
the orbital parameters $H$ and
$\tilde e^2+\tilde i^2$. 
Red ones correspond to horseshoe orbits, blue ones
to passing orbits. The solid line is the analytical 
expression (\ref{eq:my_cond}) for the boundary. 
The dashed line shows the power law
asymptote of this relation in the strongly dispersion-dominated regime.}
\label{fig:hs_separation}
\end{figure}

In the nonplanar case, on the basis of similar arguments, we suggest that
the horseshoe boundary condition in the high-velocity
regime should be replaced with 
\begin{eqnarray} 
\tilde h_{hs}=\left(\frac{k}{\tilde e^2+\tilde i^2}\right)^{1/4}.
\label{eq:my_cond_eower}
\end{eqnarray} 
To check this prediction and fix the proportionality constant $k$,
we have performed a set of orbit integrations 
 in both the shear- and dispersion-dominated 
regimes. We have separated the outcomes of encounters
into horseshoe orbits (when $H$ was changing sign as a result of
an encounter) and passing ones (when the sign of $H$ 
remained unchanged). The results are presented in 
Figure \ref{fig:hs_separation}. One can see a rather clean separation between 
the horseshoe orbits (red dots) and passing ones (blue dots). 
There are some red dots which appear in the region mostly occupied
by the passing orbits, but they originate from large-angle scattering
during close encounters and are not librating orbits as normal horseshoes are.

The separation condition (\ref{eq:my_cond_eower}) fits the boundary between
the two types of orbits very well when $\tilde e^2+\tilde i^2\gg 1$
if we take $k\simeq 8$
(dashed curve in Figure \ref{fig:hs_separation}). However, as
we move away from the strongly dispersion-dominated 
regime some deviations from
(\ref{eq:my_cond_eower}) appear. This is not surprising because
small $\tilde e$ and $\tilde i$ correspond to the shear-dominated regime,
and then the horseshoe region has a well-defined 
boundary at $\tilde h_{hs}\sim 1$. We have found that 
the shape of the horseshoe-passing boundary can be fit
rather accurately by the following
condition
\begin{eqnarray}
\tilde e^2+\tilde i^2=R^2_{hs}(\tilde h_{hs})=
k\left(\frac{1}{\tilde h_{hs}^2}-\frac{1}{b^2}\right)^2,
~~~\tilde h_{hs}<b,~~~~~~k\simeq 8,
~~~b\simeq 1.4 
\label{eq:my_cond}
\end{eqnarray}
which is shown by the solid curve on Figure \ref{fig:hs_separation}\footnote{
Our value of $b$ would predict 
$\tilde h_{hs}$ in the shear-dominated regime slightly different
from that suggested by Petit \& H\'enon (1987b):
$\simeq 1.4$ instead of $\simeq 1.2$. 
However, such a small difference is unlikely 
to be important for our purposes.}.

Using distribution function of eccentricities and inclinations 
(\ref{eq:Railey}) we can calculate the 
 fraction of passing orbits $\rho_{pass}$:
\begin{eqnarray}
\rho_{pass}(H)=\left\{
\begin{array}{l}
\exp\left[-\frac{R_{hs}^2(H)}{2\tilde\sigma_e^2}\right]+
\frac{\tilde\sigma_i^2}{\tilde\sigma_i^2-\tilde\sigma_e^2}
\left\{\exp\left[-\frac{R_{hs}^2(H)}{2\tilde\sigma_i^2}\right]-
\exp\left[-\frac{R_{hs}^2(H)}{2\tilde\sigma_e^2}\right]\right\},~~~~~~\hfill H<b,\\
1,\hfill H>b.
\end{array}
\right.
\label{eq:frac_of_eass}
\end{eqnarray}
One can see that $\rho_{pass}(H)$ changes rapidly from $0$
to $1$ between the regions of horseshoe and passing orbits (this is caused by 
the exponential dependence of $\rho_{pass}$ on $R_{hs}$ and 
the strong power law
dependence of $R_{hs}$ on $H$). This allows us
to neglect the transition region  in the dispersion-dominated 
regime and assume these different types of orbits 
to occupy different regions of the physical space. Thus we take 
passing 
 orbits to be restricted to the region $|H|>\tilde h_{hs}$ where 
$\tilde h_{hs}$ is defined such that 
\begin{eqnarray}
\rho_{pass}(\tilde h_{hs})=1/2,
\label{eq:pass_cond}
\end{eqnarray} 
and horseshoe orbits to satisfy $|H|<\tilde h_{hs}$. 
Using these conditions we can now consider passing 
orbits separately from the horseshoe ones.

%%%%%%%%%%%%%%%%%%%%%%%%%%%%%%

\subsection{Scattering on horseshoe orbits.}
\label{subsubsect:hs_scattering}

%%%%%%%%%%%%%%%%%%%%%%%%%%%%%%

The relative motion of interacting planetesimals in the horseshoe regime 
can be split into the slow shear motion of the guiding centers and the fast 
epicyclic motion. When such a separation is legitimate 
some quantities associated with the
fast motion called adiabatic invariants should be conserved
(Landau \& Lifshitz 1989).
It was shown by H\'enon \& Petit (1986) and Hasegawa \& Nakazawa (1990)
that the absolute value of the relative semimajor axis difference, 
relative eccentricity and 
relative inclination are all separately
conserved quantities in the course of a horseshoe encounter, and that they 
are the adiabatic invariants of this type of motion. 
Thus the quantities $H_0, 
{\bf \tilde e}_0, {\bf \tilde i}_0$ before the encounter
are related to their values after, $H, {\bf \tilde e}, 
{\bf \tilde i}$, as
\begin{eqnarray}
H=-H_0, ~~~{\bf \tilde e}={\bf \tilde e}_0, 
~~~{\bf \tilde i}={\bf \tilde i}_0
\label{eq:d_hs} 
\end{eqnarray}
It follows from this conservation of adiabatic invariants that
the effect of the embryo-planetesimal encounters
in the horseshoe region is to 
exchange planetesimals at symmetric orbits ($H$ and $-H$). 
Interacting bodies just 
librate  between successive close approaches when they reverse the 
direction of their motion.

Because the embryo is much more massive 
than the planetesimals and its $\tilde e$ and $\tilde i$ are 
zero, the relative motion
in the embryo-planetesimal system 
is equivalent to the motion of planetesimals. 
The scattering probability
for the horseshoe motion $\tilde P_{hs}$ looks like
$\tilde P_{hs}=
\delta(\Delta H+2H_0)
\delta(\Delta {\bf \tilde e})
\delta(\Delta {\bf \tilde i})$.
Using this result we can write down the evolution equation 
of some quantity $F$ by analogy with equations 
(\ref{eq:surf_shear})-(\ref{eq:inc_shear}) in a very simple form:
\begin{eqnarray}
\frac{\partial F(H)}{\partial t}=
\frac{|A|\mu_e^{1/3}}{\pi}|H|\left[F(-H)-F(H)\right].
\label{eq:scat_hs}
\end{eqnarray}
In our case $F$ can be the planetesimal surface density
$N$, horizontal epicyclic energy $N\tilde\sigma_e^2$ or vertical
epicyclic energy
 $N\tilde\sigma_i^2$.

%%%%%%%%%%%%%%%%%%%%%%%%%%%%%%

\subsection{Scattering on passing orbits.}
\label{subsubsect:pass_scattering}

%%%%%%%%%%%%%%%%%%%%%%%%%%%%%%

To describe the embryo-planetesimal interaction on passing orbits 
in the dispersion-dominated regime we will use the results for 
the scattering in this regime
derived in Paper II, namely equations 
(49)-(51) and (52)-(55). We substitute embryo's
surface number density in the form (\ref{eq:em_surf_density}) 
for the surface density of approaching 
bodies $N_2$, assume that $m/M_e\to 0$
 and that $\tilde \sigma_{e 2}=\tilde \sigma_{i 2}=0$. 
As a result we obtain the following system
\begin{eqnarray}
&& \frac{\partial N}{\partial t}=\frac{|A|\mu_e^{1/3}}{\pi}\left[
-\frac{\partial}{\partial H}\left(|H|\langle\Delta\tilde h\rangle
N\right)
+\frac{1}{2}
\frac{\partial^2}{\partial H^2}\left(|H|\langle(\Delta\tilde h)^2\rangle N
\right)\right],
\label{eq:em_surf}\\
&& \frac{\partial}{\partial t}\left[2N(H)\tilde \sigma_{e}^2(H)\right]=
\frac{|A|\mu_e^{1/3}}{\pi}\nonumber\\
&& \times\left[
|H|\langle\Delta ({\bf \tilde e}^2)\rangle N
-\frac{\partial}{\partial H}\left(|H| 
\langle({\bf \tilde e}^2+2{\bf \tilde e}\cdot\Delta{\bf \tilde e})
\Delta \tilde h
\rangle N\right)
+\frac{1}{2}\frac{\partial^2}{\partial H^2}\left(|H| 
\langle{\bf \tilde e}^2(\Delta \tilde h)^2
\rangle N\right)\right],
\label{eq:em_s_e}
\end{eqnarray}
and an equation for the inclination evolution analogous to (\ref{eq:em_s_e}).
These formulae are only valid in the region of space restricted by the
condition (\ref{eq:pass_cond}). Analytical expressions for the scattering 
coefficients $\langle\Delta\tilde h\rangle,\langle(\Delta\tilde h)^2,
\langle\Delta ({\bf \tilde e}^2)\rangle,
\langle({\bf \tilde e}^2+2{\bf \tilde e}\cdot\Delta{\bf \tilde e})
\Delta \tilde h
\rangle$, and 
$\langle{\bf \tilde e}^2(\Delta \tilde h)^2
\rangle$ can be found in Paper II.
The proper boundary conditions for this system
will be derived in \S \ref{subsubsect:boundary_cond}.

At this point we should address a subtle issue
which was not important for the planetesimal-planetesimal scattering
but becomes nontrivial for the embryo-planetesimal interaction.
It is related to the fact that planetesimals 
at different initial separations are driven past the embryo by the 
differential shear at different rates. 
Those which initially had $|H|\gg 1$ quickly
approach the embryo, experience scattering, and quickly depart. 
On the approach and departure stages
their orbital parameters do not have time to change except when 
they are close to the embryo.
Thus, we can assume that values of $H, {\bf \tilde e}, {\bf \tilde i}$ 
far from the embryo are the same as at the point of closest approach if
initially $|H|\gg 1$.

\begin{figure}
\vspace{14.cm}
\includegraphics{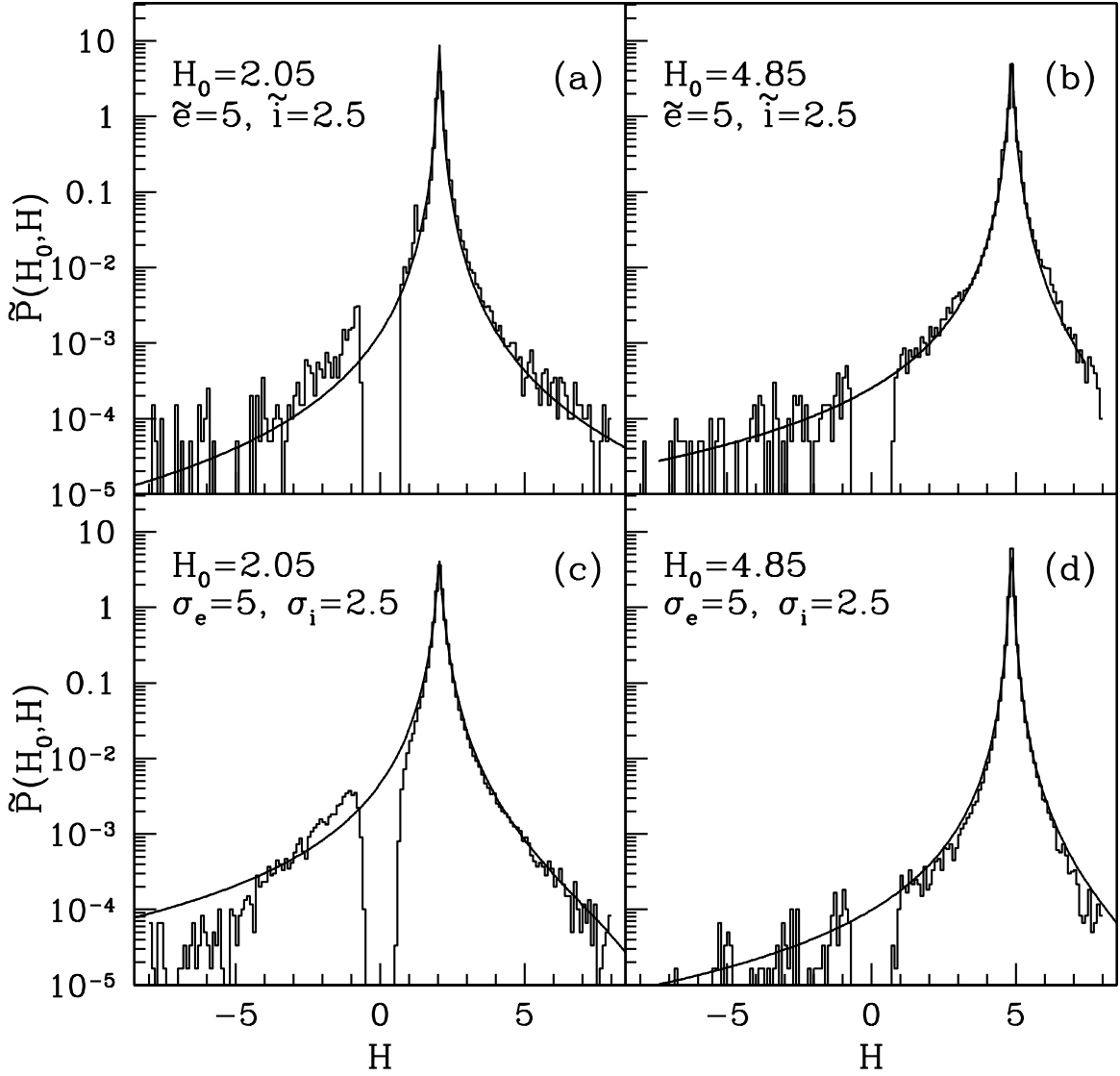}
\caption{Scattering probability $\tilde P(H_0,H)$ as a function of the final
orbital separation $H$ for a fixed initial separation $H_0$. 
The probability is calculated by binning the outcomes of orbit 
integrations ({\it histogram}) and also analytically in Appendix 
\ref{app:scat_erobab} ({\it solid line}).
Panels ({\it a,b}) display $\tilde P(H_0,H)$ for
fixed values of the initial eccentricity and inclination 
$\tilde e=2\tilde i=5$
and two values of the initial
orbital separation, $H_0=2.05$ and  $4.85$. Panels ({\it c,d}) 
display $\tilde P(H_0,H)$ for 
fixed values of the initial dispersions of eccentricity and inclination,
$\tilde \sigma_e=\tilde \sigma_i=5$.
Note the absence of scattered orbits near $H=0$ caused by the angular 
momentum exchange at large azimuthal distances (see text).}
\label{fig:Monte_Carlo_erobab}
\end{figure}

\begin{figure}
\vspace{14.cm}
\includegraphics{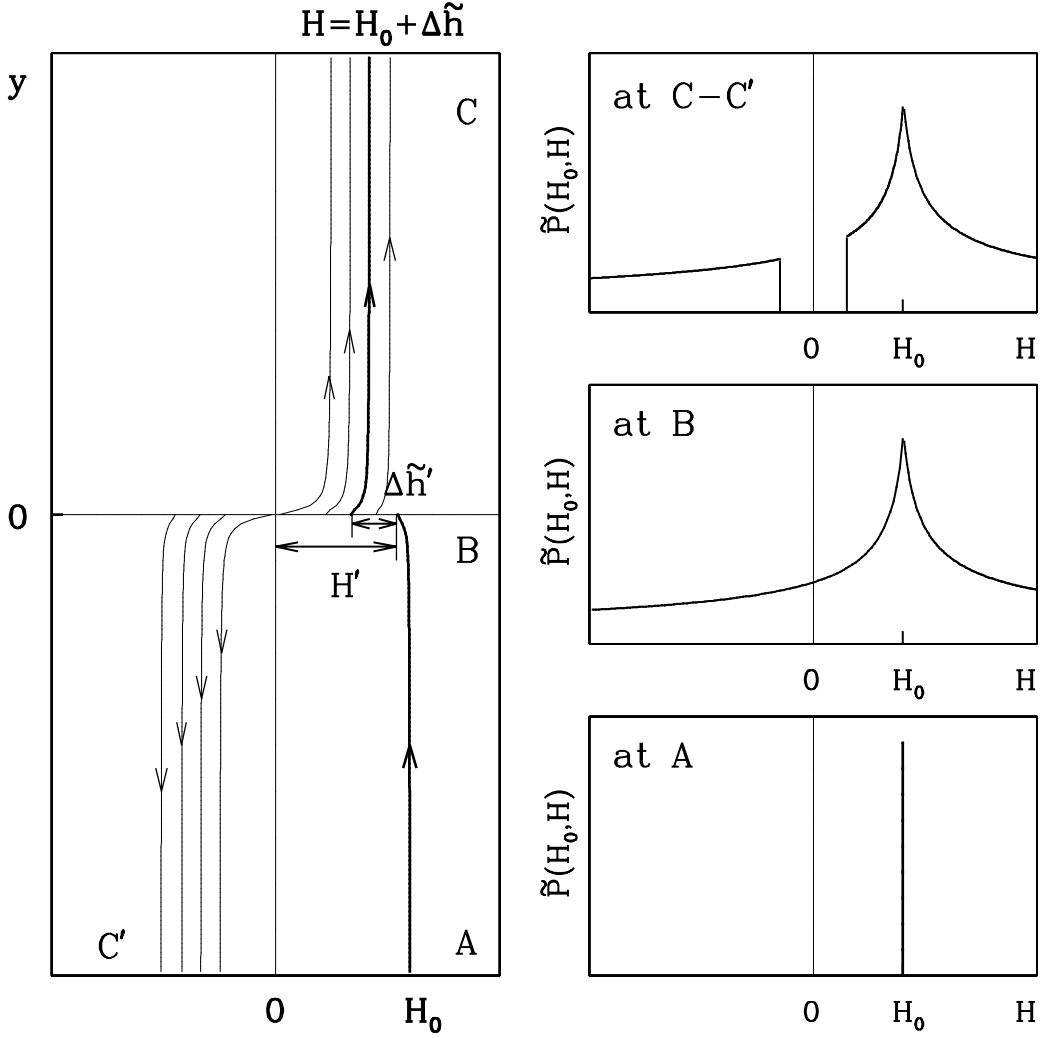}
\caption{Schematic illustration of planetesimal scattering on passing orbits
 near the 
horseshoe-passing orbits boundary. The thick solid line 
in the left panel represents the track
of the guiding center of some planetesimal initially located at $H_0$. 
As a result of angular momentum exchange with the embryo its semimajor
axis shifts to $H^\prime$ prior to scattering; then in the course of
a close encounter it is changed by $\Delta \tilde h^\prime$. 
Subsequent interaction with 
the embryo on the departure stage increases the semimajor axis
from $H^\prime+\Delta \tilde h^\prime$ to $H$.
Note that $\Delta \tilde h=H-H_0$ is not equal to 
$\Delta \tilde h^\prime$. Thin solid 
lines show the trajectories of planetesimals with the same $H_0$ but 
different epicyclic phases. The panels on the right illustrate the
origin of the gap in the probability distribution of scattering by plotting
snapshots of the semimajor axes distribution of planetesimals 
$\tilde P(H_0,H)$ 
initially (where it has $\delta$-function form, at point A), 
immediately after the encounter
(at point B), and long after (points C and C$^\prime$ coincide in the closed
box).}
\label{fig:illustration}
\end{figure}

However, planetesimals which start their motion 
not far from the horseshoe orbits boundary are not moving very
fast (because this boundary is located at $\tilde h_{hs}\le R_H$). 
As a result, for these planetesimals  the exchange of the 
angular momentum  before and after 
the close encounter can be important, just as in
the case of horseshoe orbits, where such an exchange essentially 
prevents close encounters from happening.
Consequently, the asymptotic value of the separation $H$ is not the same 
as its value at the close encounter. We must also ask whether the 
eccentricity and inclination are affected in a similar way.
In this situation ${\bf \tilde e}$ and ${\bf \tilde i}$ 
are not perfectly conserved adiabatic invariants 
(frequencies of the two superposed motions are not very strongly 
different from each other) but we probably will not make 
a huge mistake by assuming that they still are conserved 
  on the approach and departure  stages. 
Thus, we can draw the following picture of scattering
for these orbits: as the planetesimal gets closer to the embryo the absolute
value of $H$ gets smaller as a result of the angular 
momentum exchange with the embryo, while the eccentricity and inclination
are preserved (see Figure \ref{fig:illustration}). At the point of closest 
approach, the planetesimal orbital parameters experience a quick variation 
as a result of close encounter with the embryo. After that, on the departure 
stage, the embryo slowly changes the angular 
momentum of the planetesimal such that 
$|H|$ increases.

These variations of $H$ at constant ${\bf \tilde e}$ and ${\bf \tilde i}$ 
 do not contradict the conservation of the 
Jacobi constant (\ref{eq:Jacobi})
because the gravitational potential due to the embryo
changes in the course of planetesimal approach and departure as well. 
In fact, using the 
conservation of Jacobi constant we derive a prescription 
for the relation between the semimajor axis difference 
$H$ far from the embryo
and its value at the moment of the close encounter, 
$H^\prime$. To do this we note that at the moment of the 
closest approach the distance between the embryo and planetesimal 
in units of the embryo's Hill radius 
is of order $\sqrt{\tilde e^2+\tilde i^2}$. Then the conservation
of the Jacobi constant  (\ref{eq:Jacobi}) tells us that
$H^2-(H^\prime)^2\sim(\tilde e^2+\tilde i^2)^{-1/2}$. 
This empirical expression may be inaccurate if one applies it
for $\tilde e, \tilde i\le 1$. 
This can be remedied  using the following approximate prescription
\begin{eqnarray}
(H^\prime)^2=H^2-
\frac{c}{\sqrt{\tilde e^2+\tilde i^2+d^2}},~~~~~c\simeq 1.8,~~~~d\simeq 2,
\label{eq:h_hprime}
\end{eqnarray}
where the numerical values of constants $c$ and $d$ were fixed 
by comparison with 
 orbit integrations. Whenever the planetesimal disk has a distribution of 
eccentricities and inclinations we replace $\tilde e$ and $\tilde i$
in (\ref{eq:h_hprime}) with $\tilde \sigma_e$ and $\tilde \sigma_i$.
The relationship (\ref{eq:h_hprime}) is 
certainly not very accurate. 
However, this transformation is 
satisfactory for our purposes since the complete separation 
of horseshoe and passing orbits 
we are assuming is a rather crude approximation anyway.

To better understand the consequences and importance
of this effect
it is instructive to look at the distribution function of planetesimal
scattering $\tilde P(H_0,H)$ in the dispersion-dominated regime
[by definition $\tilde P(H_0,H)dH$ is the probability that a planetesimal 
initially at $H_0$ is scattered into the interval $(H,H+dH)$]. We
calculate this function analytically in Appendix \ref{app:scat_erobab}  
and represent it in the following form: 
\begin{eqnarray}
\tilde P(H_0,H)=\frac{1}{2\ln\Lambda}
\frac{\langle (\Delta \tilde h)^2\rangle
+\langle \Delta \tilde h\rangle\Delta H}
{(|\Delta H|+d)^3},~~~\Delta H=H-H_0,
\label{eq:dif_erobab_cutoff1}
\end{eqnarray}
where
\begin{eqnarray}
d\approx \left[\frac{\langle (\Delta \tilde h)^2\rangle}
{2\ln\Lambda}\right]^{1/2},
\label{eq:d_def1}
\end{eqnarray}
$\langle \Delta \tilde h\rangle$ and $\langle (\Delta \tilde h)^2\rangle$
are the scattering coefficients (see Paper II),
and $\ln\Lambda$ is the usual Coulomb logarithm. To compute $\tilde P(H_0,H)$ 
accounting for the distribution of planetesimal eccentricities 
and inclinations one simply needs to use the values of 
$\langle \Delta \tilde h\rangle$ and 
$\langle (\Delta \tilde h)^2\rangle$ 
averaged over this distribution (they are given in \S 4 of Paper II for
the case of Gaussian distribution of $\tilde e$ and $\tilde i$).

One can also obtain this distribution function from numerical orbit
integrations. To do this we perform Monte-Carlo simulations  
in two different ways: by fixing the absolute values 
of the eccentricity and inclination of the planetesimals 
(only their epicyclic phases are chosen randomly), and by picking their
orbital elements from the distribution (\ref{eq:Railey})
with fixed dispersions.
In both cases the initial difference of semimajor axes $H_0$ is the same
for each simulation. In Figure \ref{fig:Monte_Carlo_erobab}a,b
we plot $\tilde P(H_0,H)$  for fixed $\tilde e=2\tilde i=5$ and in 
Figure \ref{fig:Monte_Carlo_erobab}c,d
we do this for fixed $\tilde \sigma_e=2\tilde \sigma_i=5$. The analytical 
distribution given by (\ref{eq:dif_erobab_cutoff1}) is also plotted for 
each case. One can see that it generally agrees very well with the 
numerical results. The large variations of $\tilde P(H_0,H)$ in the 
outer wings of the simulated
distributions are caused by statistical noise. 
When $|H-H_0|\le 1$ the agreement between theory
and simulations is quite remarkable even for the distributions
averaged over $\tilde e, \tilde i$ (Figure \ref{fig:Monte_Carlo_erobab}c,d). 

However, one can also immediately notice one important feature 
of the simulated $\tilde P(H_0,H)$: there are 
no orbits scattered in a region with 
a width $\sim R_H$ around the embryo's location, and this is in contrast 
with the analytical result (\ref{eq:dif_erobab_cutoff1}) which 
exhibits no such feature. A similar ``gap'' in the distribution of scattered 
orbits can also be noticed in the numerical calculations performed 
by Greenzweig \& Lissauer (1990) and Ohtsuki \& Tanaka (2002). 

From our previous discussion the reason for the appearance of this
gap becomes
quite apparent. It is the gravitational interaction of the particles on
passing orbits located close to the horseshoe region
with the embryo which drives them away from its
orbit, corresponding to $H=0$. 
If we were to take the same probability distributions not far away
from the embryo in the azimuthal direction but immediately after the 
scattering (within $\sim R_H$ from the embryo along the azimuthal direction)
we would not find such a ``gap''. 
However, as the differential shear slowly 
increases the azimuthal separation of the interacting bodies 
their mutual angular
momentum exchange leads to a gradual increase of $|H|$ similar 
to that happening on the horseshoe orbits. As a result, a 
conspicuous ``gap'' appears in $\tilde P(H_0,H)$ near $H=0$. 
This process is illustrated in Figure \ref{fig:illustration}.

This effect has important consequences for the calculation of scattering
on passing orbits with radial separations $\sim R_H$ (or $|H|\sim 1$).
The difference between  $H$ far from the embryo
(which we take as one of the planetesimal coordinates) and
its value at the moment of the close encounter introduces 
changes into the computation of scattering coefficients.
Indeed, if before and after the encounter the values of planetesimal 
semimajor axis are $H_0$ and $H$, and at the point 
of closest approach it has the value of $H^\prime$, we can write that
(see Figure \ref{fig:illustration})
\begin{eqnarray}
H^\prime(H_0)+\Delta 
\tilde h^\prime[H^\prime(H_0)]=
H^\prime(H_0+\Delta \tilde h),
\label{eq:relationship}
\end{eqnarray}
where $\Delta \tilde h^\prime$ and $\Delta \tilde h$ are changes of
$H^\prime$ and $H$ correspondingly. Our calculation of
scattering coefficients in \S 4 of Paper II gives
us only the value of $\Delta \tilde h^\prime$ (change of the planetesimal
semimajor axis at the closest approach). 
Here however we
are interested in $\Delta \tilde h=H-H_0$. To relate them we assume that both
changes are small which is always a good approximation in the 
dispersion-dominated regime. Then, expanding the r.h.s. of 
(\ref{eq:relationship}) up to the second order in $\Delta \tilde h$
we find that
\begin{eqnarray}
\Delta \tilde h(H_0)=\left(\frac{\partial H^\prime}
{\partial H_0}\right)^{-1}
\Delta \tilde h^\prime-\frac{1}{2}
\frac{\partial^2 H^\prime}{\partial H_0^2}
\left(\frac{\partial H^\prime}
{\partial H_0}\right)^{-3}
(\Delta \tilde h^\prime)^2+O((\Delta \tilde h^\prime)^3),\nonumber\\
(\Delta \tilde h(H_0))^2=\left(\frac{\partial H^\prime}
{\partial H_0}\right)^{-2}
(\Delta \tilde h^\prime)^2+O((\Delta \tilde h^\prime)^3).
\label{eq:relat}
\end{eqnarray} 
At the same time our assumption of approximate adiabatic conservation of
$\tilde {\bf e}$ and $\tilde {\bf i}$ implies that the changes of 
these quantities evaluated
far from the embryo are the same as they are in the course of an encounter 
near the embryo. It also means that in calculating scattering coefficients
we should use the values of eccentricity and inclination 
characterizing the planetesimal orbit
far from the embryo. 

The relationships represented by equations (\ref{eq:relat}) are utilized 
when we calculate scattering coefficients in equations 
(\ref{eq:em_surf})-(\ref{eq:em_s_e}).

%@@@@@@@@@@@

\subsubsection{Boundary conditions for the scattering on passing orbits.}
\label{subsubsect:boundary_cond}

%@@@@@@@@@@@

The simplest way to derive the boundary conditions for equations 
(\ref{eq:em_surf})-(\ref{eq:em_s_e}) is to consider the Fokker-Planck
equation for the evolution of the distribution function of planetesimals
$f({\bf \Gamma})$ caused by embryo-planetesimal encounters. Here
${\bf \Gamma}$ is a set of orbital elements characterizing the planetesimal 
orbital state: $\Gamma_1=H, (\Gamma_2,\Gamma_3)={\bf \tilde e}$ 
(vector eccentricity)
and  $(\Gamma_4,\Gamma_5)={\bf \tilde i}$ (vector inclination).
Following conventional wisdom
(Lifshitz \& Pitaevskii 1981; Binney \& Tremaine 1987) 
we can write the evolution equation of $f$ caused by weak encounters:
\begin{eqnarray}
\frac{\partial f}{\partial t}=-2|A|\left[\sum\limits_{i=1}^{5}
\frac{\partial }{\partial \Gamma_i}\left(f|H|\langle\Delta \Gamma_i\rangle
\right)+
\frac{1}{2}
\sum\limits_{i=1}^{5}
\frac{\partial^2}{\partial \Gamma_i\partial  \Gamma_j}
\left(f|H|\langle\Delta  \Gamma_i\Delta \Gamma_j\rangle\right)\right]
=-\frac{\partial F_i}{\partial \Gamma_i},
\label{eq:FP_em__master}
\end{eqnarray}
where the flux in the $i$-th direction is given by
\begin{eqnarray}
F_i=2|A|\left[f|H|\langle\Delta \Gamma_i\rangle-\frac{1}{2}
\sum\limits_{i=1}^{5}
\frac{\partial}{\partial \Gamma_j}
\left(f|H|\langle\Delta \Gamma_i\Delta \Gamma_j\rangle\right)\right].
\label{eq:FP_flux}
\end{eqnarray}
Averaging of quantities like $\langle\Delta \Gamma_i\rangle$ 
is performed here only over the possible outcomes of scattering  
and not over $d\tilde {\bf e}d\tilde {\bf i}$. 
The factor $2|A||H|$ takes care of the linear
shear velocity in the planetesimal disk (changes of various quantities 
caused by the scattering are assumed to be per 
encounter and not per unit of time).
Note that we could have obtained equations 
(\ref{eq:em_surf})-(\ref{eq:em_s_e}) directly from 
(\ref{eq:FP_em__master}).

Particles on passing orbits do not
penetrate into the region of horseshoe orbits (as a result of
our presumed complete separation of these two types of orbits).
Thus the component of the flux $F_H$ must vanish at the boundaries 
$H=\tilde h_{hs}$ and $H=-\tilde h_{hs}$:
\begin{eqnarray}
&& f|H|\langle\Delta H\rangle-\frac{1}{2}
\frac{\partial}{\partial ({\bf \tilde e}^2)}
\left[f|H|\langle\Delta H\Delta({\bf \tilde e}^2)\rangle\right]
\nonumber\\  
&& 
  -\frac{1}{2}
\frac{\partial}{\partial ({\bf \tilde i}^2)}\left[f|H|\langle\Delta H
\Delta({\bf \tilde i}^2)\rangle\right]-\frac{1}{2}
\frac{\partial}{\partial H}\left[f|H|\langle(\Delta H)^2\rangle\right]=0
~~~
\mbox{at}~~H=\pm\tilde h_{hs}.
\label{eq:flux}
\end{eqnarray}

Our next step is to multiply condition (\ref{eq:flux}) by
$d(\tilde {\bf e}^2)d(\tilde {\bf i}^2)$, $\tilde {\bf e}^2d
(\tilde {\bf e}^2)d(\tilde {\bf i}^2)$ 
and $\tilde {\bf i}^2d(\tilde {\bf e}^2)d(\tilde {\bf i}^2)$ and integrate it
over ${\bf \tilde e}^2$-${\bf \tilde i}^2$ space. We obtain
 as a result (integrating by parts where needed) that
\begin{eqnarray}
&& N|H|\langle\Delta \tilde h\rangle-\frac{1}{2}
\frac{\partial}{\partial H}
\left[N|H|\langle(\Delta \tilde h)^2\rangle\right]=0,
\label{eq:surf_cond}\\
&& N|H|\langle{\bf \tilde e}^2\Delta \tilde h\rangle-\frac{1}{2}
\frac{\partial}{\partial H}\left[N|H|\langle 
{\bf \tilde e}^2(\Delta \tilde h)^2\rangle\right]
+\frac{1}{2}N|H|\langle\Delta({\bf e}^2)\Delta \tilde h\rangle=0,
\label{eq:s_e_cond}\\
&& N|H|\langle{\bf \tilde i}^2\Delta \tilde h\rangle-\frac{1}{2}
\frac{\partial}{\partial H}\left[N|H|\langle 
{\bf \tilde i}^2(\Delta \tilde h)^2\rangle\right]
+\frac{1}{2}N|H|\langle\Delta({\bf \tilde i}^2)\Delta \tilde h\rangle=0~~~
\mbox{at}~~H=\pm\tilde h_{hs},
\label{eq:s_i_cond}
\end{eqnarray}
where now $\langle...\rangle$ means integration not only over the probability
distribution of scattering but also over the 
$ \tilde {\bf e}^2$-$\tilde{\bf i}^2$ space (and we use the
more familiar notation $\langle\Delta \tilde h\rangle$ instead of 
$\langle\Delta H\rangle$, etc.).
In getting these results we have taken into account that 
 all scattering coefficients vanish
at the boundaries of the velocity space (which are at
infinity). 

As we discussed in Paper II terms like
$\langle(\Delta {\bf \tilde e})^2\Delta \tilde h\rangle$  
are third order in perturbations and can be neglected.
Thus we can approximate $\langle\Delta({\bf \tilde e}^2)\Delta \tilde 
h\rangle$ with
$2\langle({\bf \tilde e}\cdot\Delta{\bf \tilde e})\Delta \tilde h\rangle$.
As a result we find that
\begin{eqnarray}
&& N|H|\langle\Delta \tilde h\rangle-\frac{1}{2}
\frac{\partial}{\partial H}\left[N|H|\langle(\Delta \tilde h)^2
\rangle\right]=0,
\label{eq:surf_cond_reduced}\\
&& N|H|\langle ({\bf \tilde e}^2+{\bf \tilde e}
\cdot\Delta {\bf \tilde e})\Delta \tilde h\rangle-\frac{1}{2}
\frac{\partial}{\partial H}\left[N|H|\langle 
{\bf \tilde e}^2(\Delta \tilde h)^2\rangle\right]=0,
\label{eq:s_e_cond_reduced}\\
&& N|H|\langle ({\bf \tilde i}^2+{\bf \tilde i}
\cdot\Delta {\bf \tilde i})\Delta \tilde h\rangle-\frac{1}{2}
\frac{\partial}{\partial H}\left[N|H|\langle 
{\bf \tilde i}^2(\Delta \tilde h)^2\rangle\right]=0~~~
\mbox{at}~~H=\pm\tilde h_{hs}.
\label{eq:s_i_cond_reduced}
\end{eqnarray}
Note the specific combination ${\bf \tilde e}^2+{\bf \tilde e}
\cdot\Delta {\bf \tilde e}$, not
${\bf \tilde e}^2+2{\bf \tilde e}\cdot\Delta {\bf \tilde e}$ as in 
equation (\ref{eq:em_s_e}) ---
this is important for the conservation of the Jacobi constant.

Do equations (\ref{eq:em_surf})-(\ref{eq:em_s_e}) conserve the 
integrated Jacobi constant (\ref{eq:int_Jacobi}) and 
the total number of planetesimals (\ref{eq:int_number})?
If we integrate (\ref{eq:em_surf}) over $dH$ (which would give
us the total number of planetesimals in the region of interest), we
find that condition 
(\ref{eq:surf_cond_reduced}) ensures the conservation of
$N^{tot}$. The same is true for the integrated Jacobi
constant: when we substitute equations  (\ref{eq:em_surf})-(\ref{eq:em_s_e})
into the definition (\ref{eq:int_Jacobi}) we find (after some cumbersome but
straightforward calculations) that conditions 
(\ref{eq:surf_cond_reduced})-(\ref{eq:s_i_cond_reduced}) 
guarantee the conservation of $J^{tot}$ up to the second order in 
the perturbed quantities. 
Thus, expressions (\ref{eq:surf_cond_reduced})-(\ref{eq:s_i_cond_reduced}) 
provide  a desired set of self-consistent boundary conditions for 
the equations (\ref{eq:em_surf})-(\ref{eq:em_s_e}).

%@@@@@@@@@@@

\subsubsection{Comparison with numerical orbit integrations.}
\label{subsubsect:orbit_integrations}

%@@@@@@@@@@@

To check our predictions about the behavior of the planetesimal disk 
properties derived in the previous sections we have performed a set 
of numerical simulations. We have integrated the orbits of 
test particles, starting at large 
azimuthal separation from the embryo, 
and observe the changes of their orbital
parameters as they experience gravitational interactions with
the planetary embryo. Initial orbital parameters are chosen 
randomly from the distribution 
of eccentricities and inclinations (\ref{eq:Railey}). 
The semimajor axes of particles are assumed to be uniformly
distributed within some radial interval around the embryo. The width of 
this interval is large enough that boundary effects are not important.
We typically calculate about $2\times 10^5$ different orbits to reduce 
statistical noise. 
Each orbit experiences several hundred passages past the embryo during which
its orbital elements change. Between the conjunctions we randomize the 
epicyclic phases of the planetesimals to mimic the effect of 
planetesimal-planetesimal interaction which is assumed to destroy 
any resonances in the system (the hypothesis of molecular chaos). However the 
absolute values of eccentricity and inclination are not affected by 
this procedure. In the course of the integration embryo is assumed to be 
immobile.

The number of consecutive passages is 
dictated by the condition that the system evolves for at least one dynamical 
time within region of phase space for which close encounters can occur 
 $|H|\le \tilde \sigma_e$. 
In the shear-dominated regime this corresponds to a single passage
at radial separation $\sim R_H$
(i.e. dynamical time $t_{dyn}\sim t_{syn}$ --- synodic period at $H=1$)
because scattering is strong in this case. At the same time equations 
(\ref{eq:em_surf})-(\ref{eq:em_s_e}) and analytical expressions for 
the scattering
coefficients in Paper II show that in the dispersion-dominated regime the
dynamical time $t_{dyn}\sim t_{syn}\tilde \sigma_e^5$ when 
$\tilde \sigma_e\sim\tilde \sigma_i$. This timescale
 can easily take several hundred orbital 
passages which require that one follow planetesimal orbital evolution for 
rather a long time.

\begin{figure}
\vspace{15cm}
\includegraphics{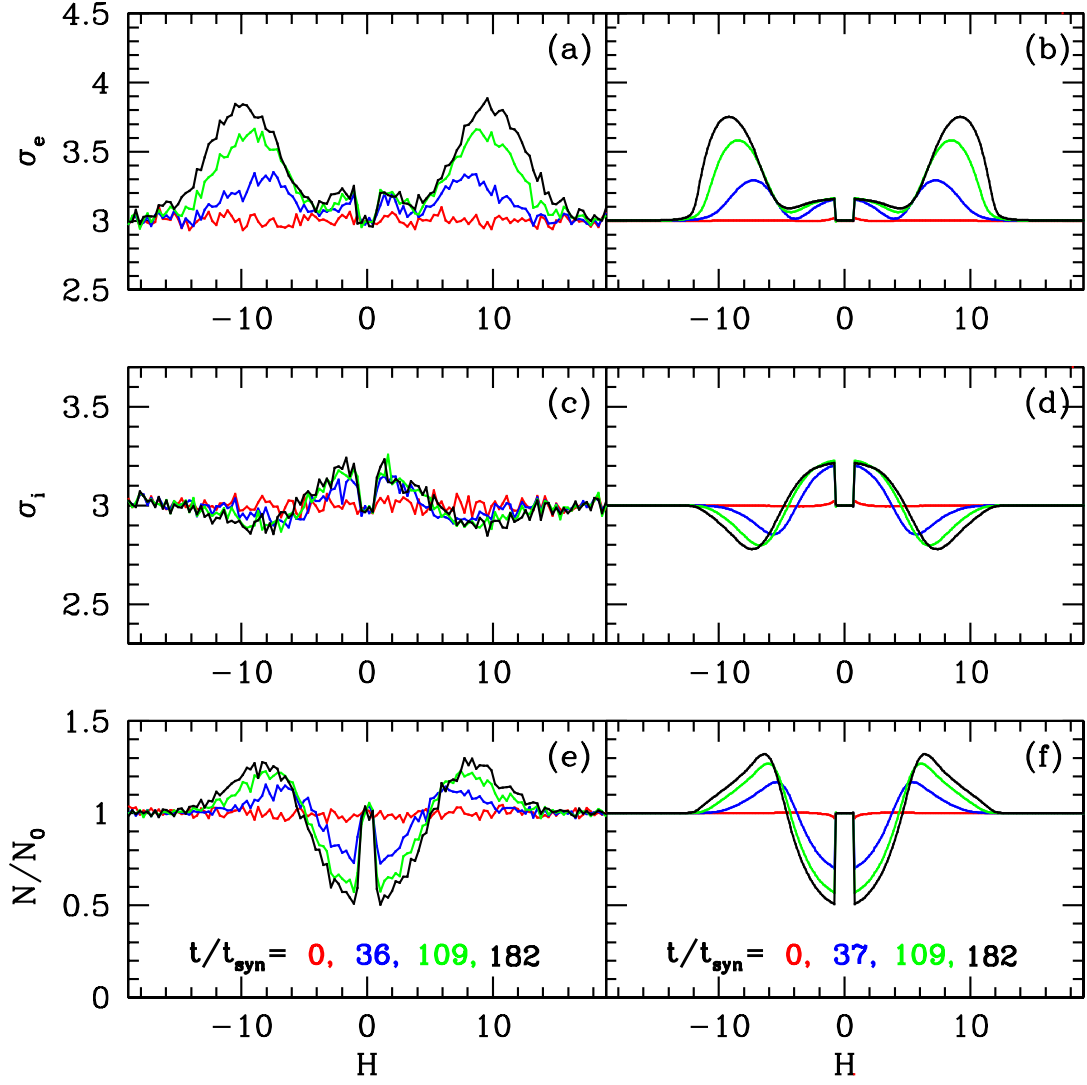}
\caption{Evolution of the planetesimal disk properties driven by the presence
of a massive protoplanetary embryo. Initially 
$\tilde \sigma_e=\tilde \sigma_i=3$. 
The plots contain numerical {\it (left row)} and analytical 
{\it (right row)} time sequences of profiles of $\tilde \sigma_e$ ({\it a,b}), 
$\tilde \sigma_i$ ({\it c,d}), and dimensionless surface density normalized by
its value at infinity ({\it e,f}). Curves of different colors
represent profiles measured at specific moments of time normalized by the
synodic period at a separation of $H=1$ (which are shown 
in panels ({\it e}) and ({\it f}) by corresponding color coding; they are 
slightly different for numerical and analytical curves).}
\label{fig:evolution_3_3}
\end{figure}

\begin{figure}
\vspace{15cm}
\includegraphics{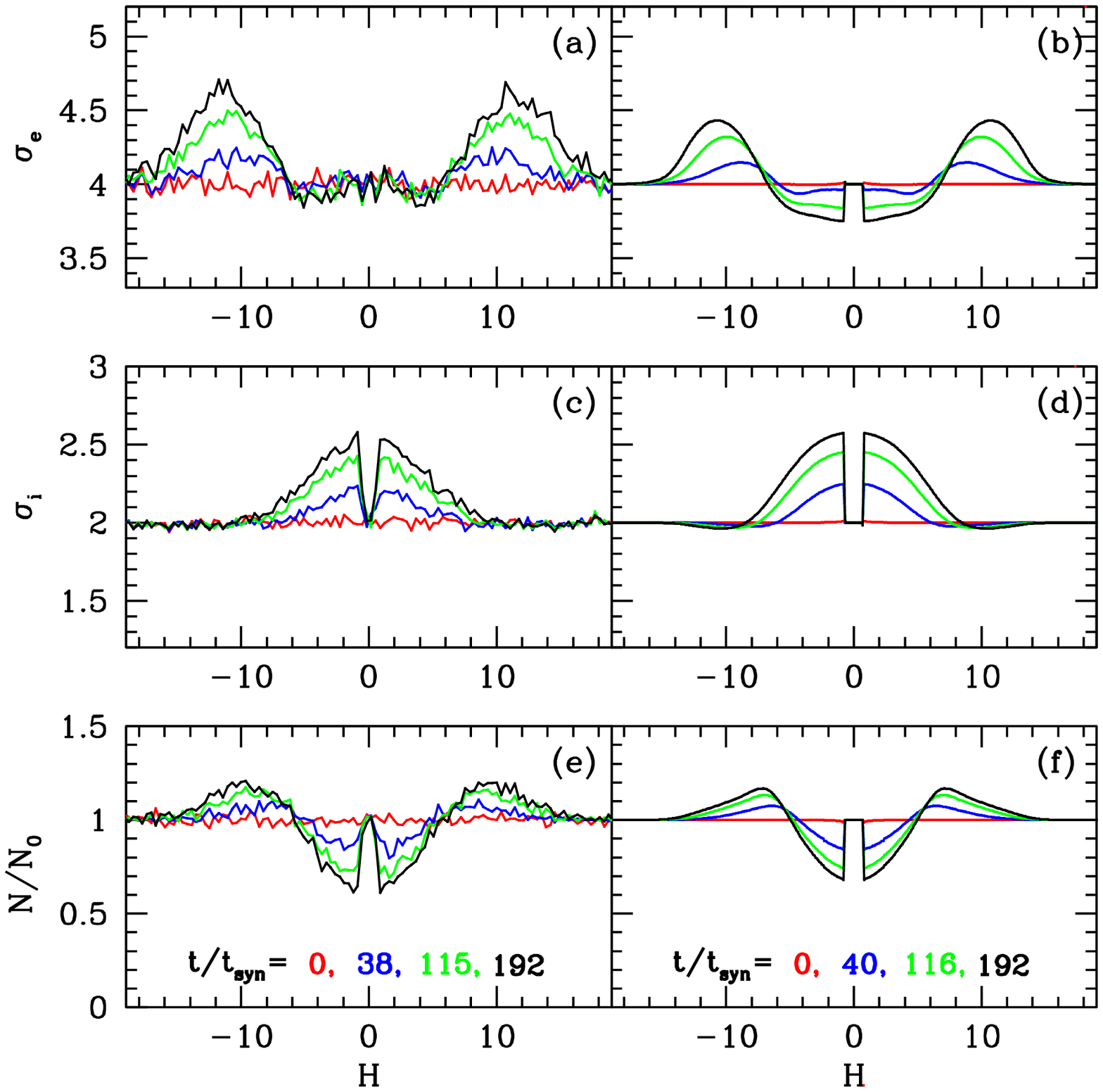}
\caption{The same as Figure \ref{fig:evolution_3_3} but for initial 
$\tilde \sigma_e=4, \tilde \sigma_i=2$.}
\label{fig:evolution_4_2}
\end{figure}

In Appendix \ref{app:intermediate_velocity} we
describe our treatment of the intermediate (between shear- and 
dispersion-dominated) velocity regime and make comparisons with 
the results of orbit integrations. Here
we perform this procedure for the dispersion-dominated regime.  
In Figures \ref{fig:evolution_3_3} and \ref{fig:evolution_4_2} 
we display time sequences of 
profiles of horizontal and vertical velocity dispersions 
[panels ({\it a,b}) and ({\it c,d}) correspondingly], 
and dimensionless surface density normalized by
its value at infinity [panels ({\it e,f})]. We show both
the results of orbit integrations (left row of panels) and 
our analytical predictions (right row of panels).
Disk evolution in the region of the horseshoe orbits is described
using equation (\ref{eq:scat_hs}). In the region of passing orbits
[outside of the interval $(-\tilde h_{hs}, \tilde h_{hs})$, see 
\S \ref{subsubsect:hs_separation}] disk evolution is governed by
partial differential evolution equations 
(\ref{eq:em_surf})-(\ref{eq:em_s_e}). We solve them using fully
implicit scheme (Press \etal 1988) with the boundary conditions
(\ref{eq:surf_cond_reduced})-(\ref{eq:s_i_cond_reduced}) imposed
at $H=-\tilde h_{hs}$ and $H=\tilde h_{hs}$ and scattering 
coefficients computed  analytically in Paper II.
The conversion (\ref{eq:h_hprime}) is used throughout the calculation 
and the boundary of the horseshoe region is described by 
formula (\ref{eq:pass_cond}). For the factor $\Lambda$ entering
the Coulomb logarithm in the 
expressions for the scattering coefficients we use the following 
prescription\footnote{To avoid problems at $\tilde \sigma_e,
\tilde\sigma_i\le 1$ we use instead of $\ln\Lambda$ the more accurate
expression $(1/2)\ln(1+\Lambda^2)$ (see Binney \& Tremaine 1987).}:
\begin{eqnarray}
\Lambda=\tilde\sigma_i(2\tilde\sigma_e^2+\tilde\sigma_i^2).
\label{eq:lam_choice}
\end{eqnarray}
Constant coefficients in this formula are roughly fixed using comparison with 
orbit integrations (but our final results depend on them very weakly).

Figure \ref{fig:evolution_3_3} shows the disk evolution
in the case when the initial planetesimal 
dispersions of eccentricity and inclination 
are both equal to $3$. One can see an excellent 
qualitative agreement between
the results of orbit integrations and analytical theory. All the features 
of the spatial distributions of disk quantities are well reproduced by
the solutions of the analytical equations. 
There is also a reasonable degree of 
quantitative agreement between them although there are also some 
minor differences: the analytical equations predict somewhat faster evolution
of $\tilde \sigma_i$ and a slightly smaller radial extent
of the excited region than orbit integrations do (this might be caused by
the effect of the distant encounters which theory 
can not reproduce very well, see Appendix \ref{app:intermediate_velocity}). 
In Figure \ref{fig:evolution_4_2} we display the same comparison of 
numerical and analytical results but now for the case of 
$\tilde\sigma_e=4, \tilde\sigma_i=2$. This corresponds to the ratio
$\tilde\sigma_i/\tilde\sigma_e=0.5$ which is within the range 
$0.45-0.55$ thought to be realized 
in homogeneous Keplerian planetesimal disks (Ida \etal 1993; 
Stewart \& Ida 2000). One can see again a very good consistency between
the two approaches, especially for the evolution of $N$ and $\tilde\sigma_i$. 
There is a slight discrepancy between the predictions of two methods
 for the evolution of $\tilde\sigma_e$ near the
horseshoe-passing orbits boundary. However, since equations 
(\ref{eq:em_surf})-(\ref{eq:em_s_e}) preserve the Jacobi constant one 
can be sure that the evolution of the surface density (which is 
driven by the redistribution of the angular momentum in the planetesimal 
disk) is always consistent
with the evolution of epicyclic energy independent of the details
of the spatial distributions of various quantities.

Planetesimals with the semimajor axes close to the embryo's orbit 
perform horseshoe motion (see \S \ref{subsect:em_dispersion}). 
As a result, surface density of guiding centers 
in this region stays constant in time as can be seen in 
Figures \ref{fig:evolution_3_3} and \ref{fig:evolution_4_2}. 
Note that our orbit integrations 
do not exhibit the concentration of planetesimals at the embryo's 
semimajor axis which is present in numerical 
calculations of Tanaka \& Ida (1997).
In their case this effect might have been 
caused by the analytical simplifications which
were employed in Tanaka \& Ida (1996, 1997) to speed up  
orbit integrations. 

Figures \ref{fig:evolution_3_3} and \ref{fig:evolution_4_2} show 
that the embryo does not clear a complete gap on a dynamical timescale 
in the dispersion-dominated regime, although it does produce  a significant 
depression in the surface density. 
Note that we plot the surface density of guiding centers of 
planetesimals $N$ in Figures \ref{fig:evolution_3_3} and 
\ref{fig:evolution_4_2} rather than the instanteneous surface density. The
depression formed in the instanteneous surface density at the location
of the embryo is weaker than the gap in $N$ because 
epicyclic motion allows planetesimals to penetrate inside the depression
of $N$.  The absence of a clear gap is due to the weakness of the
individual embryo-planetesimal scattering in the dispersion-dominated 
regime. In the shear-dominated
disk, scattering is much stronger and a gap is cleared by the embryo
on a rather short timescale (see Paper I and Appendix 
\ref{app:intermediate_velocity}). It is also obvious that the evolution
of the disk surface density is accompanied by a considerable change in 
kinematic properties of planetesimal population, as required by
the conservation of Jacobi constant. Thus, the  dynamical 
evolution of the disk is an important ingredient of the gap formation 
process which justifies the need for the self-consistent theory such 
as the one presented here.

 The agreement between the analytical and numerical results 
in the dispersion-dominated regime is in general
pretty good given the approximate nature
of the theory and the small number of fitting parameters we are 
using --- essentially only the constants in 
(\ref{eq:my_cond}), (\ref{eq:h_hprime}), and (\ref{eq:lam_choice}) are free
parameters, and  it turns out that the solutions of equations 
(\ref{eq:em_surf})-(\ref{eq:em_s_e}) depend on their particular choice 
only very weakly. We believe that minor discrepancies between the outcomes
of analytical and numerical approaches can 
be accounted for by going to the next order in $1/\ln\Lambda$ 
and properly including the effects of the distant encounters. 
We expect our analytical formulation to be even more accurate for
larger values of $\tilde \sigma_e$, $\tilde \sigma_i$ but we have not 
investigated these because the computational requirements of the
numerical simulations become prohibitive.
We postpone the detailed exploration of these subjects for the future.

%@@@@@@@@@@@

\section{Embryo's accretion rate.}
\label{sect:accretion}

%@@@@@@@@@@@

The accretion rate of planetesimals by the embryo $\dot M$
is another observable
(in addition to the spatial distributions of $N$, $\tilde\sigma_e$, and 
$\tilde\sigma_i$) which can be computed both analytically and 
numerically and used as a check of our calculations. In addition, it is 
a very important quantity by itself for any realistic modelling of 
planet formation. A lot of work has been devoted to the study of
the planetary accretion
in homogeneous planetesimal disks in the context of the planetary mass
growth (Greenzweig \& Lissauer 1990, 1992) and the 
origin of planetary spins and obliquities (Dones \& Tremaine 1993). 
The results of these authors are not directly applicable to the 
inhomogeneous disks that are our primary focus  
but can be used as limiting cases to check our
analytical predictions in the more general case of nonuniform distribution
of planetesimals.

In Appendix \ref{app:accr_rate} we calculate $\dot M$
in different velocity regimes. In the dispersion-dominated case
such a computation is made possible by the use of the two-body approximation
and we find that
\begin{eqnarray}
\dot M=m\frac{\Omega R_e^2}{8 a_e^2}\int\limits_{-\infty}^\infty 
N(H)\frac{|H|dH}{\tilde \sigma_e^2\tilde \sigma_i^2}
e^{-(H^\prime)^2/(2\tilde \sigma_e^2)}
\left[\frac{(H^\prime)^2}{4} U_+(H^\prime,\tilde \sigma_e,\tilde \sigma_i)+
\frac{2}{p}U_-(H^\prime,\tilde \sigma_e,\tilde \sigma_i)\right],
\label{eq:accr}
\end{eqnarray}
where
\begin{eqnarray}
U_\pm=\int\limits_0^\infty dr(1+r)^{\pm \frac{1}{2}}
e^{-\frac{1}{2}\left(\alpha_e^2+\alpha_i^2\right)r}
I_0\left[\frac{1}{2}\left(\alpha_i^2-\alpha_e^2\right)r\right],~~~
\alpha_e=\frac{H^\prime/\tilde\sigma_e}{2\sqrt{2}},~~~
\alpha_i=\frac{H^\prime/\tilde\sigma_i}{2\sqrt{2}},
\label{eq:Upm_def}
\end{eqnarray}
$R_e$ is the embryo's physical radius, $a_e$ is its semimajor axis, 
$p=R_e/R_H$, $I_0$ is a modified Bessel function of order zero, 
and $H^\prime(H)$ is
given by (\ref{eq:h_hprime}).
Integration over $dH$ in (\ref{eq:accr})
should exclude the region $(-\tilde h_{hs},\tilde h_{hs})$ corresponding 
to the horseshoe orbits.

In the
shear-dominated regime we use simple scaling arguments to fix the accretion
rate dependence on the physical variables of the system and find that
\begin{eqnarray}
\dot M\approx 5N^{inst}(H_{coll})\frac{ m\Omega R_e R_H}
{\langle\tilde\sigma_i\rangle a_e^2}
\label{eq:accr_shear}
\end{eqnarray}
where $N^{inst}(H_{coll})$ is the surface density of 
planetesimals on orbits leading to collisions 
with the embryo (at $H=H_{coll}\approx 1.4$,
see Appendix \ref{app:accr_rate}), defined
by equation (\ref{eq:N_inst}),
and $\langle\tilde\sigma_i\rangle$ is a measure of 
$\tilde\sigma_i$ in the inhomogeneous disk obtained by
averaging $\tilde\sigma_i$ over the interval 
$(-2R_H,2R_H)$ [from where most of the 
planetesimals get accreted in the shear-dominated regime, see Petit \& H\'enon
(1986)].
In Appendix
\ref{app:intermediate_velocity} we demonstrate how to use these results to
cover the regime of intermediate velocity dispersion.

\begin{figure}[t]
\vspace{11.5cm}
\includegraphics{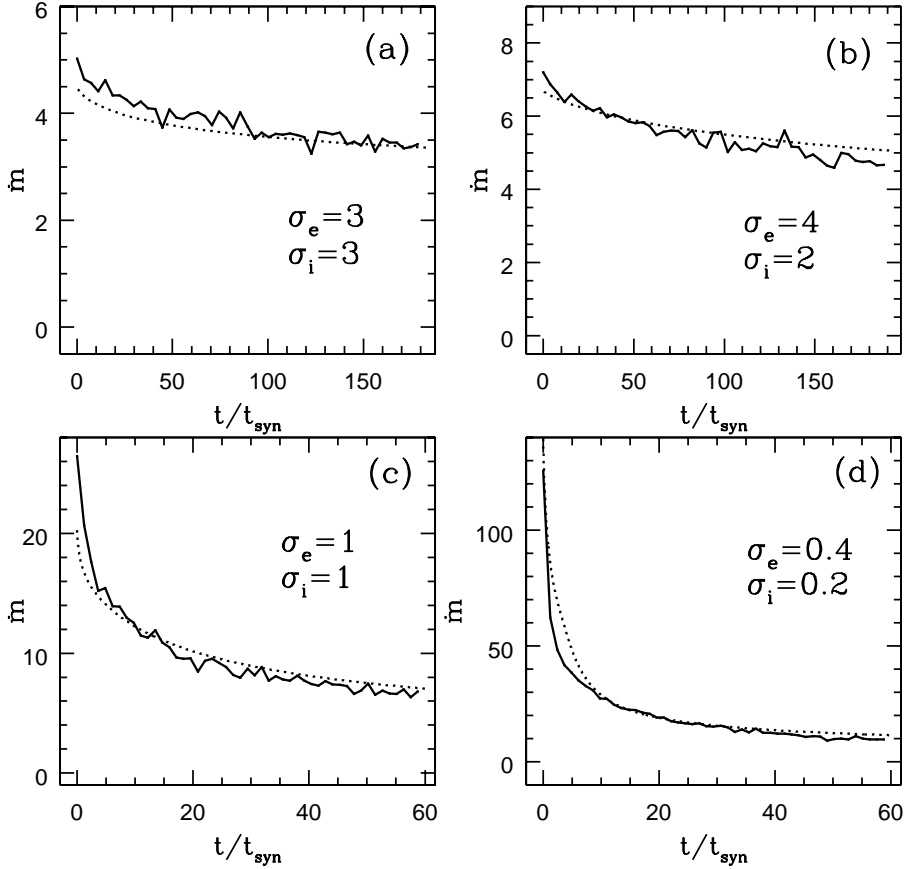}
\caption{Dimensionless accretion rate $\dot m=\dot M/(\Omega R_e^2 \Sigma_0)$
($R_e$ is the radius of the planet, $\Sigma_0$ is the 
surface mass density of the disk far from the embryo) as a function of time 
(expressed in synodic periods $t_{syn}$ at the separation of $1~R_H$)
for different values of $\tilde \sigma_e$ and $\tilde \sigma_i$. 
Results of orbit integrations are shown by thick curve, and the dotted line 
shows analytical predictions.
Accretion rates for the dispersion-dominated regime  are presented in 
panels $(a)$ and $(b)$, while panels $(c)$ and $(d)$ display $\dot m$
for intermediate velocity regime.}
\label{fig:acc_rate}
\end{figure}
 
Figure \ref{fig:acc_rate} shows the comparison of our analytical 
predictions with the results of orbit integrations. We plot the 
dimensionless accretion rate $\dot m=\dot M/(\Omega R_e^2 \Sigma_0)$
($R_e$ is the radius of the planet, $\Sigma_0=mN(\infty)/a_e^2$ is the 
dimensional 
surface mass density of the disk far from the embryo) as a function
of time normalized by the synodic period $t_{syn}$ corresponding
to a separation of $1~R_H$ between the orbits of the embryo and planetesimal.
Numerical accretion rates were computed during the calculation
of the spatial and temporal evolution of various disk quantities described 
in \S \ref{subsubsect:orbit_integrations}. They are displayed 
 by the thick solid line in Figure \ref{fig:acc_rate}. 
Analytical results [using
an interpolation given by formulae (\ref{eq:m_acc_interp}) and 
(\ref{eq:Pi_def}) where needed] are shown by a dotted line.  
One can see that the agreement
is reasonably good even in the intermediate velocity
regime\footnote{Note that the specific form of interpolation function 
(\ref{eq:Pi_def}) was chosen to fit several sets of $\tilde\sigma_e$ and
$\tilde\sigma_e$, not only those displayed in Figure \ref{fig:acc_rate}. 
In all
studied cases numerical and analytical results agree within $(10-30)\%$.} 
(Figure \ref{fig:acc_rate}c,d). The theoretical 
results in the dispersion-dominated
regime (Figure \ref{fig:acc_rate}a,b) do not depend on the interpolation 
and their good agreement with the numerical accretion rates confirms the 
validity of the analytical approach developed in Appendix 
\ref{app:accr_rate}. We will be using this prescription for the embryo's 
accretion rate in upcoming work when dealing with the evolution of the 
planetesimal disk coupled to the embryo's growth.

%@@@@@@@@@@@

\section{Direction of the embryo-planetesimal interaction.}
\label{sect:repulsion}

%@@@@@@@@@@@

From the results of the orbit integrations 
described in \S \ref{subsubsect:orbit_integrations}
one can see that the embryo-planetesimal interaction
leads to a decrease in the surface density of planetesimals
near the embryo in the region of passing orbits. The width of the 
surface density depression is
typically of the order of the planetesimal eccentricity dispersion 
$\tilde\sigma_e$. At the same time the distribution of  
planetesimals on horseshoe orbits stays unaffected by the embryo.

This is very similar to the behavior of the surface density 
in the shear-dominated regime, as
studied in Paper I. In that case the reason for such behavior was very clear: 
scattering of a planetesimal initially on a circular orbit can only 
increase its eccentricity and inclination which leads to the increase
of $|H|$ (a result of Jacobi constant conservation) and, 
consequently, to the repulsion of planetesimal orbits. In Paper I
planetesimals were kept
on circular orbits at all times, and a gap with a width of several $R_H$
was carved near the embryo's orbit.

In the kinematically hot regime the reasoning
is not so simple. From equation (\ref{eq:dif_erobab_cutoff1}) and Figure
\ref{fig:Monte_Carlo_erobab} one can see that scattered planetesimals 
can have $\Delta |H|<0$ as well as $\Delta |H|>0$ for a fixed $H$
depending on their eccentricities, inclinations, and epicyclic phases. 
Moreover, from our calculation of the scattering coefficients in
the dispersion-dominated regime (Paper II, see also the analogous result
in Ida \etal 2000) we know that 
$\langle\Delta \tilde h\rangle<0$. Using this result  
Ida \etal (2000) suggested that in the 
kinematically hot planetesimal disk embryo should {\it attract} planetesimal 
orbits rather than {\it repel} them. The outcome of such an interaction 
would be a growth of planetesimal
surface density at the embryo's location. 

However, this conclusions is misleading. 
What really determines the behavior of the planetesimal surface 
density near the 
embryo and the nature of the embryo-planetesimal interaction is 
the direction of the {\it flux} of planetesimals. The embryo 
repels planetesimals and carves a gap 
when this flux is directed away from the embryo;
it attracts planetesimals and increases their nearby surface density 
when this flux is directed towards the embryo.
The condition  $\langle\Delta \tilde h\rangle<0$ by itself
cannot guarantee that the planetesimal flux is directed towards 
the embryo's orbit because these quantities are not simply 
related\footnote{One can imagine a scattering probability
function which transports most of the planetesimals away from the embryo
shifting them by a small $\Delta H>0$; 
at the same time a small fraction of planetesimals can be 
scattered towards the embryo and have large negative values of $\Delta H$.
With properly chosen weights one can always ensure that 
$\langle\Delta \tilde h\rangle<0$ while the bulk of material moves
away from the embryo forming a gap around its orbit.} . 
Another complication precluding the use of the sign of 
$\langle\Delta \tilde h\rangle$ as an indicator of planetesimal disk
evolution is that $\Delta H<0$ (for  $H>0$) does not always mean that 
planetesimal orbit gets closer to the embryo. If for example $\Delta H<-2H$
then the absolute value of the semimajor axis separation after the encounter 
is larger than it was before, which is indicative of repulsion  rather
than attraction.

\begin{figure}[t]
\vspace{9.5cm}
\includegraphics{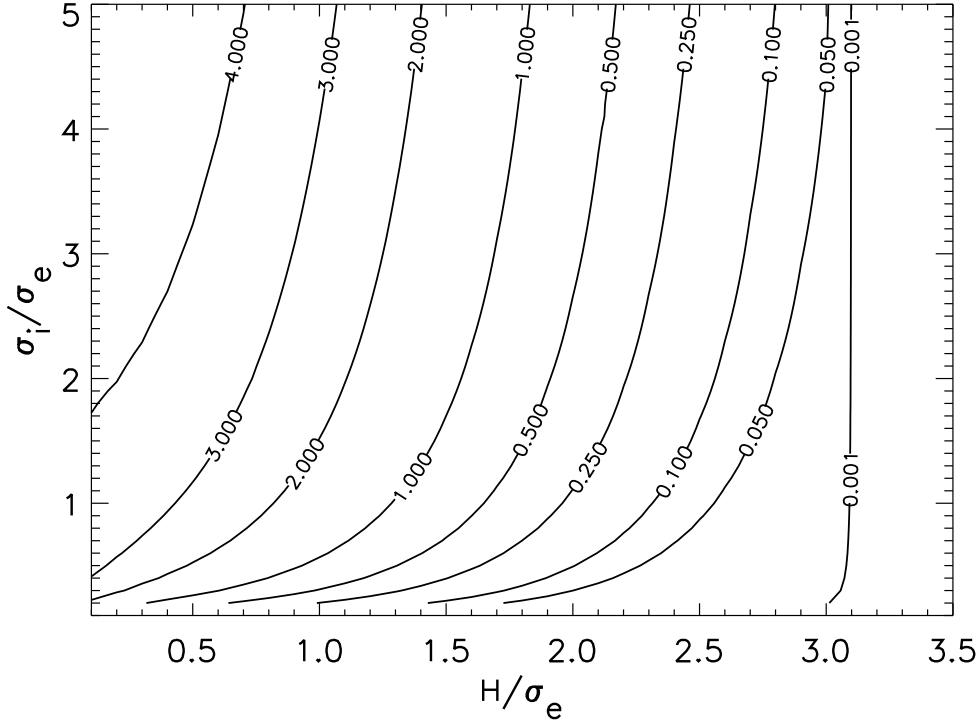}
\caption{Contour plot of the dimensionless planetesimal flux  
$\pi {\cal F}/(|A|\mu_e^{1/3})$ as a function of 
$\tilde \sigma_i/\tilde \sigma_e$ --- ratio of the vertical to horizontal 
velocity dispersions, and $H/\tilde \sigma_e$ --- radial distance from the
embryo scaled by the RMS epicyclic radius. Note that this flux is always
positive for positive $H$ meaning that the embryo {\it repels} planetesimals.
The flux is an odd function of $H$.}
\label{fig:flux}
\end{figure}

Using the method described in Appendix A of Paper I, 
we can calculate the outward flux of 
planetesimals ${\cal F}$ driven by the embryo-planetesimal interactions.
We evaluate this flux in a homogeneous disk.
For the general scattering probability function $\tilde P(H_0,\Delta H)$
[$\tilde P(H_0,\Delta H)d\Delta H$ is the probability for 
a planetesimal initially at $H_0$ to have $\Delta H$ in the 
range $(\Delta H,\Delta H+d\Delta H)$] this flux 
can be written as (see Papers I and II)
\begin{eqnarray}
{\cal F}=\frac{|A|\mu_e^{1/3}}{\pi}\left[
\int\limits_0^\infty d(\Delta H)\int\limits_{H-\Delta H}^H
dH_0|H_0|\tilde P(H_0,\Delta H)-\int\limits^0_{-\infty} d(\Delta H)
\int\limits^{H-\Delta H}_H dH_0|H_0|\tilde P(H_0,\Delta H)\right].
\label{eq:flux1}
\end{eqnarray}
Differentiating this expression w.r.t. $H$ gives us
\begin{eqnarray}
\frac{\partial {\cal F}}{\partial H}=\frac{|A|\mu_e^{1/3}}{\pi}
\left[|H|-\int\limits^\infty_{-\infty} d(\Delta H)
|H-\Delta H|\tilde P(H-\Delta H,\Delta H)\right].
\label{eq:flux2}
\end{eqnarray}
Since the planetesimal flux in a homogeneous disk 
has to vanish far from the embryo we can
integrate (\ref{eq:flux2}) to obtain
\begin{eqnarray}
{\cal F}=\frac{|A|\mu_e^{1/3}}{\pi}
\int_{-\infty}^H dH_1\left[|H_1|-\int\limits^\infty_{-\infty} d(\Delta H)
|H_1-\Delta H|\tilde P(H_1-\Delta H,\Delta H)\right].
\label{eq:flux3}
\end{eqnarray}

This expression [or (\ref{eq:flux1})] is valid for an arbitrary function
$\tilde P(H_0,\Delta H)$. In principle we can use this
general form but it is more convenient to apply one of the most important 
properties of the interaction in the dispersion-dominated regime: 
that scattering is dominated by weak encounters, meaning that we
can expand in (\ref{eq:flux3}) the integral in brackets in a Taylor series
in $\Delta H$.
As a result, after some intermediate transformations we find that
planetesimal flux in the initially homogeneous disk is given by
\begin{eqnarray}
{\cal F}(H)=\frac{|A|\mu_e^{1/3}}{\pi}
\left[|H|\langle\Delta \tilde h\rangle-\frac{1}{2}
\frac{\partial }{\partial H}
\left(|H|\langle(\Delta \tilde h)^2\rangle\right)\right].
\label{eq:flux_final}
\end{eqnarray}
This expression could have also been obtained directly 
from equation (\ref{eq:em_surf}) or (\ref{eq:FP_flux}). It is
clear from the equation (\ref{eq:flux_final}) why the sign of 
$\langle\Delta \tilde h\rangle$ cannot tell us the direction of the 
planetesimal evolution: the presence of additional term with the
derivative of 
$|H|\langle(\Delta \tilde h)^2\rangle$ in r.h.s. of (\ref{eq:flux_final})
which cannot in general be related\footnote{A relation 
between the first and second order diffusion 
coefficients exists if a homogeneous distribution is a steady-state 
solution of the Fokker-Planck equation
see Lifshitz \& Pitaevskii (1981). In our case this is not 
true --- embryo's gravity tends to make disk inhomogeneous 
by carving out a gap.} to $\langle\Delta \tilde h\rangle$
means that there is no direct
relation between the signs of ${\cal F}(H)$ and 
$\langle\Delta \tilde h\rangle$.

Using the analytical formulae for $\langle\Delta \tilde h\rangle$ and
$\langle(\Delta \tilde h)^2\rangle$ derived in Paper II, we can 
calculate this flux in the Fokker-Planck approximation as a function
of distance from the embryo and  the ratio of the vertical to horizontal
velocity dispersions. The result of such a calculation is presented
in Figure \ref{fig:flux}. The shape of the contours is universal in the 
coordinates $H/\tilde \sigma_e, \tilde \sigma_i/\tilde \sigma_e$, i.e. it
does not depend on the embryo mass, planetesimal mass, or 
other parameters. In plotting 
${\cal F}$ we have also neglected the 
effect of interaction at large azimuthal 
distances near the horseshoe region boundary 
 described in \S \ref{subsubsect:pass_scattering} 
(it can be easily taken into account
and will only slightly shift the contours near $H\sim 1$.) 
One can easily see from Figure \ref{fig:flux} that ${\cal F}>0$ when 
$H>0$ for all values of $H/\tilde \sigma_e$ and 
$\tilde \sigma_i/\tilde \sigma_e$ which we 
explored (and ${\cal F}<0$ for $H<0$ by symmetry). 
This implies that the embryo tends to {\it repel} planetesimals by driving
a flux directed away from its orbit. This conclusion disagrees  with 
the predictions based on the sign of $\langle\Delta \tilde h\rangle$ 
(Ida \etal 2000). The planetesimal flux vanishes as 
$H/\tilde \sigma_e$ grows 
but distant encounters (not included in Figure \ref{fig:flux})
will dominate at large $H/\tilde \sigma_e$ and they also always
drive planetesimals away from the embryo (Goldreich \& Tremaine 1982;
Petit \& H\'enon 1987a). Thus, it appears that repulsion is the only 
possible outcome of the embryo-planetesimal gravitational interaction even
in the dispersion-dominated velocity regime.

Figure \ref{fig:flux} seems to imply that the planetesimal surface density
will always grow because ${\cal F}$ decreases with $H$. This is not the 
case because the boundary condition (\ref{eq:surf_cond_reduced})
forces ${\cal F}=0$ at the horseshoe-passing orbit boundary 
$H=\tilde h_{hs}$.  This means that there is
actually a transition zone near $H=\pm\tilde h_{hs}$ in which 
flux steeply increases from
$0$ to its value given by (\ref{eq:flux_final}). The width of this 
zone is infinitesimal initially but it rapidly expands as the planetesimal 
disk evolves under the action of the embryo's gravity. Within the transition 
region ${\cal F}$ is positive and increases with $H$ 
meaning that the planetesimal surface density decays there. 
This is exactly what was observed in the analytical calculations
and numerical orbit integrations in \S \ref{subsubsect:orbit_integrations}.
The slowdown of the disk evolution obvious from Figures 
\ref{fig:evolution_3_3} and \ref{fig:evolution_4_2} is caused not
only by the growth of epicyclic energy of the disk (which lengthens the disk
dynamical timescale) but also by the gradual expansion of the transition 
zone leading to a decrease in the planetesimal flux gradient and
slower disk evolution.

%*************************************************************
%*************************************************************

\section{Discussion and conclusions.}
\label{sect:conclusion}

%*************************************************************
%*************************************************************

We have studied the embryo-planetesimal 
interaction in the gravitational field of a central star. Two different cases 
were explored: when the interaction between the embryo and planetesimals occurs
in the shear-dominated  and in the
dispersion-dominated regimes.
The treatment of the first case parallels that of Paper I
but is complicated by the fact that now we explicitly include the evolution of
not only the surface density but also the eccentricity and 
inclination of planetesimals in the disk.

Our study of the dispersion-dominated regime relies on the methods   
of kinetic theory, and it uses many of the results obtained in 
Paper II. However our present treatment
is more refined since the description of embryo-planetesimal scattering
requires clarifying many details which were not important for the 
planetesimal-planetesimal interactions. In particular we have to study 
not only passing but also horseshoe orbits of planetesimals to determine 
the spatial distribution of the disk properties. To do this we propose 
a condition which separates the horseshoe and passing orbits and check 
its viability using numerical orbit integrations. 
Angular momentum exchange between the embryo and 
planetesimal long before and
after their closest approach turns out to be important for
the scattering on passing orbits near the horseshoe boundary. 
We illustrate this point by comparing
the analytical scattering probability function with the one obtained
from numerical integrations. A simple method to  account for this effect in
our Fokker-Planck approach is proposed. 
%We have also included the effect
%of the embryo on the planetesimals separated from it by large distance ---
%distant encounters. 
Taken together all these refinements are 
shown to provide rather good agreement with the results of the 
numerical orbit integrations. Thus we hope to have grasped the most
important features of the embryo-planetesimal interaction by our 
theoretical approach.

This does not mean that our treatment of the embryo-planetesimal 
interaction is complete. We have only focussed on 
the most important, dominant effects, and there is certainly room
for additional refinements, which would farther improve the agreement with 
numerical results.  The calculation of the scattering coefficients
to the next order in $1/\ln\Lambda$ would provide us with 
subdominant corrections which might be important for modest values of 
$\tilde\sigma_e$ and $\tilde\sigma_i$. 
One can certainly do a better job in treating distant encounters, 
calculating various coefficients 
entering formulae (\ref{eq:my_cond}), (\ref{eq:h_hprime}), etc. or
the interpolating functions of 
Appendix \ref{app:intermediate_velocity},
using a larger set of numerical orbit integrations.
%\footnote{This is a natural
%route along which, for example, the study of homogeneous planetesimal disk
%dynamics proceeded in the last decade, see Stewart \& Wetherill (1988),
%Ida (1990), Ida \& Makino (1992), Stewart \& Ida (2000).}. 
On a somewhat deeper level,
one can try to come up with a more sophisticated 
treatment of the horseshoe-passing orbits separation
(instead of the complete spatial separation
of these two types of orbits assumed in this paper). 
Our purely deterministic treatment 
of the shear-dominated
regime can also be improved, which would ameliorate the comparison with 
numerical results in the intermediate velocity regime 
(see Appendix \ref{app:intermediate_velocity}).
The recoil of the embryo and the excitation of the embryo's eccentricity 
and inclination in the course
of the planetesimal gravitational scattering can be important in some 
applications, such as the embryo's migration (Tanaka \& Ida 1999) or 
its interaction with the embryos nearby (Tanaka \& Ida 1997). 
Our treatment relies on the use of the Schwarzschild
velocity distribution which was demonstrated to be applicable in
the dispersion-dominated regime (Greenzweig \& Lissauer 1992), but 
the deviations
from this assumption could be important e.g. in the intermediate velocity 
regime, and this subject can also be pursued farther.
All these refinements would better the quantitative agreement between
the analytical theory and numerical results. 
But reasonably good accord is provided
even by our basic treatment developed here, especially in the 
dispersion-dominated case where the assumptions we make are the 
most justifiable.

We also dwell upon the question of the direction 
of the embryo-planetesimal
interaction, namely whether it leads to the repulsion of planetesimal orbits
from the embryo or to their attraction. The latter outcome has been favored 
in some scenarios (Ida \etal 2000) and is based on the fact that
in the dispersion-dominated regime embryo {\it on average}
tends to attract planetesimal semimajor axes toward its orbit. We demonstrate, 
however, that the average change of planetesimal semimajor axes 
cannot serve a standard for determining the
direction of the embryo-planetesimal interaction because the transport of
the angular momentum (associated with the changes in semimajor
axes) is not the same as the bulk motion of the disk material.
We propose our own criterion for judging the embryo-planetesimal scattering
outcome, and show that the embryo always {\it repels} planetesimals
rather than attracts them in an initially homogeneous disk thereby 
carving out
a gap in the distribution of the planetesimal semimajor axes
[which is in contrast to claims by Ida \etal (2000)].  Our own 
numerical results support this conclusion.

The combination of our results for planetesimal-planetesimal gravitational
scattering presented in Paper II and the theory for the embryo-planetesimal
interaction developed here now allows us to study 
the planetesimal disk dynamics 
with these processes operating simultaneously. 
It also provides a framework to which various refinements 
and additional physical mechanisms can be naturally added. 
Our results are not restricted in applications to the problem of the 
formation of planetary systems but can also be used for studying their present
day evolution, e.g. the dynamics of asteroid and Kuiper belts affected
by massive bodies inside or near them.
Our results for the accretion
rate of massive body immersed in inhomogeneous planetesimal disk 
(\S \ref{sect:accretion} \& Appendix \ref{app:accr_rate}) 
allow us also to include self-consistently
the embryo's 
mass growth into consideration. We will examine protoplanetary disk evolution 
with all these effects working together in a future work (Rafikov 2002b).

\acknowledgements

I am grateful to my advisor, Scott Tremaine, for his patient guidance 
and valuable advice. The financial support of this work by 
 the Charlotte Elizabeth Procter Fellowship and NASA grant NAG5-10456
is thankfully acknowledged.

\appendix

%*************************************************************
%*************************************************************

\section{Scattering probability function in the dispersion-dominated regime.}
\label{app:scat_erobab}

%*************************************************************
%*************************************************************

In this appendix we calculate the scattering distribution function
$\tilde P(H,\Delta H)$ in the dispersion-dominated regime. 
This function is defined such that $\tilde P(H,\Delta  H)
d(\Delta H)$ is the probability that two bodies initially separated
by $H$ will have a change of $H$ in the interval
$(\Delta H,\Delta H+d(\Delta H))$ during an encounter. 
This probability will initially be considered as a function of
also $\tilde e$ and $\tilde i$. Then we will average it over the distribution
of $\tilde e$ and $\tilde i$ and obtain the averaged value of
$\tilde P(H,\Delta H)$ as a function of
$\tilde \sigma_e$ and $\tilde \sigma_i$. We will extensively use the 
results and notation of \S 4 of Paper II which should be consulted for 
further details and clarifications.

To do this we will employ the two-body
approximation. We first relate $\Delta H$ to the change of the
relative velocity in the $y$-direction $\Delta v_y$ using formula 
$\Delta H=2\Delta v_y/(\Omega R_H)$. 
Then we switch to the coordinates $l$ 
(impact parameter) and $\phi$
 (angle in the plane perpendicular to the
direction of approach velocity\footnote{The origin from
which this angle is counted is not important here.}) 
to characterize the trajectory of the
bodies. Using these coordinates one can express $\Delta v_y$ as
(Binney \& Tremaine 1987)
\begin{eqnarray}
\Delta v_y=-\Delta v_\parallel\frac{v_y}{v}+\Delta v_\perp
\frac{\sqrt{v_x^2+v_z^2}}{v}\cos\phi,
\label{eq:dvy}
\end{eqnarray}
where $v_x,v_y,v_z$ and $v$ are different components of the planetesimal 
approach velocity and its magnitude given by equations (58) and (60) 
of Paper II; expressions for $\Delta v_\parallel$ and $\Delta v_\perp$ 
are given by (67) {\it ibid}. This allows us to express $\Delta H$ as
\begin{eqnarray}
\Delta H=\frac{2}{1+\left(l/l_0\right)^2}\left[
-H+2\left(l/l_0\right)\tilde v_1 \cos\phi\right].
\label{eq:del-h}
\end{eqnarray}
where $\tilde v_1^2=\tilde e^2+\tilde i^2-H^2$ 
($\tilde v_1^2$ is always positive because $|H|<\tilde e$ for the kind
of scattering we consider here)
and
$l_0=G(m_1+m_2)/v^2=R_H (\Omega r_H/v)^2$ (when minimum approach distance 
between interacting bodies $\le l_0$ large-angle two-body scattering  
takes place).

For a fixed $\Delta H$ this equation can be rewritten
as an equation for $u=(l/l_0)\cos\phi$ and  $w=(l/l_0)\sin\phi$:
\begin{equation}
\left(u-\frac{2\tilde v_1}{\Delta H}\right)^2+w^2=R^2,~~~~~~
R^2=\frac{(\Delta H-\Delta H_-)
(\Delta H_+-\Delta H)}{\Delta H^2},
\label{eq:l_sol}
\end{equation}
where
\begin{eqnarray}
\Delta H_-=-H-
\sqrt{H^2+4 \tilde v_1^2},~~~
\Delta H_+=-H+\sqrt{H^2+4 \tilde v_1^2}.
\label{eq:dh_em}
\end{eqnarray}
Clearly this is only possible if 
\begin{eqnarray}
\Delta H_-<\Delta H <\Delta H_+.
\label{eq:condit}
\end{eqnarray}
Thus, the difference in planetesimal semimajor axes cannot 
increase by more than $\Delta H_+$ or decrease by more than
$\Delta H_-$ during the scattering. Equation (\ref{eq:l_sol}) is
simply the equation of a shifted circle  with radius $R$ in coordinates 
$u,w$.

To calculate the probability distribution function 
$\tilde P(H,\Delta H)$ 
we use the following identity:
\begin{eqnarray}
\tilde P(H,\Delta H)=
\left\{
\begin{array}{l}
\frac{d}{dz}\tilde P(H,\Delta H^\prime<z)\Big|_{z=\Delta H},
~~~~~~~~~~\Delta H<0,\\
-\frac{d}{dz}\tilde P(H,\Delta H^\prime>z)\Big|_{z=\Delta H},
~~~~~~~~\Delta H>0,
\end{array}
\right.
\label{eq:splitting}
\end{eqnarray}
where $\tilde P(H,\Delta H^\prime<z)$ and 
$\tilde P(H,\Delta H^\prime>z)$ are complementary cumulative 
probability distributions. We calculate them in the following way.

It is clear that the probability 
$\tilde P(H,{\bf S})$
of some outcome of scattering
is equal to the ratio of the area of the region  
of the space of epicyclic phases $\tau-\omega$ corresponding to 
the outcome manifold ${\bf S}$
and $4\pi^2$.
Using the conversion between the integration over $\tau-\omega$ and
$l-\phi$ represented by equation (61) of Paper II we can write this as
\begin{eqnarray}
\tilde P(H,{\bf S})=\int\limits_{\bf S}
\frac{d\tau d\omega}{4\pi^2}=
\frac{4}{3\pi}\frac{1}{R_H^2}
\frac{v}{\Omega R_H}\frac{1}
{\tilde i|H|\sqrt{\tilde e^2-H^2}}
\int\limits_{\bf S}ldl\frac{d\phi}{2\pi}.
\label{eq:cumul}
\end{eqnarray}
Thus we can always transform our problem into the calculation of the 
area of region ${\bf S}$ in $l-\phi$ coordinates. Here we are interested
in the area of the region given by $\Delta H<z$ for $z<0$ and
by $\Delta H>z$ for $z>0$.

One can easily see that both conditions correspond to the inner part 
of a circle given by equation (\ref{eq:l_sol}). Thus the 
area of the region in $l-\phi$ phase space bounded by these conditions
is always represented by a single formula (no matter whether $z$ is 
positive or negative):
\begin{eqnarray}
\int\limits_{\bf S}ldl\frac{d\phi}{2\pi}=\frac{\pi R^2}{2\pi}=
\frac{l_0^2}{2}
\frac{(z-\Delta H_-)
(\Delta H_+-z)}{z^2}.
\label{eq:l_esi_surf}
\end{eqnarray}
Substituting this result into (\ref{eq:cumul}) and then into
(\ref{eq:splitting}) we obtain (using the definitions of $v$ and $l_0$) that
\begin{eqnarray}
\tilde P(H,\Delta H)=\frac{4}{3\pi}
\frac{4\tilde v_1^2-H\Delta H}{|\Delta H|^3}
\frac{1}
{\tilde i|H|\sqrt{\tilde e^2-H^2}
\left[\tilde e^2+\tilde i^2-(3/4)H^2\right]^{3/2}}
\label{eq:differ}
\end{eqnarray}
if $\Delta H_-<\Delta H<\Delta H_+$, and 
$\tilde P(H,\Delta H)=0$ otherwise.
Recalling the definition of $\tilde v_1^2$ and using the 
definitions of scattering
coefficients $\langle \Delta \tilde h\rangle_{\tau,\omega}$ and
$\langle (\Delta \tilde h)^2\rangle_{\tau,\omega}$ 
given by equations 
(73), (74), and (77) of Paper II
we can rewrite (\ref{eq:differ})
in a more appealing form:
\begin{eqnarray}
\tilde P(H,\tilde e,\tilde i,\Delta H)=\frac{1}{2\ln\Lambda}
\frac{\langle (\Delta \tilde h)^2\rangle_{\tau,\omega}
+\langle \Delta \tilde h\rangle_{\tau,\omega}\Delta H}
{|\Delta H|^3},
\label{eq:dif_erobab}
\end{eqnarray}
where the dependence on $\tilde e$ and $\tilde i$ is hidden in 
$\langle \Delta \tilde h\rangle_{\tau,\omega}$ and
$\langle (\Delta \tilde h)^2\rangle_{\tau,\omega}$.

The expression for the differential probability 
given by equation (\ref{eq:dif_erobab}) is most accurate for 
large angle scattering
(i.e. for $\Delta H\ga 1$)  because
this means very small impact parameters implying that the two-body scattering 
assumption is very good. This allows one to use this probability function
to study strong scattering in the Hill approximation. 
On the other hand when $\Delta H\to 0$ 
the expression (\ref{eq:dif_erobab})
 clearly diverges. This 
divergence is unphysical because the two-body assumption
becomes very bad there. Small $\Delta H$ means weak
scattering which only occurs for  
large impact parameters $l$. In real planetesimal
disks $l$ is always restricted to be $\la \tilde i R_H$. 
Thus our formula (\ref{eq:dif_erobab}) is only applicable for
$|\Delta H|\ga \Delta H_{min}$, where we find 
$\Delta H_{min}$ by setting
$l\approx \tilde i R_H\gg l_0$ in equation (\ref{eq:del-h}):
\begin{eqnarray}
\Delta H_{min}\approx\frac{2\tilde v_1}{\tilde i}\left(\frac{\Omega R_H}
{v}\right)^2\ll 1.
\label{eq:dhmin}
\end{eqnarray}

When $|\Delta H|\la \Delta H_{min}$ the scaling provided
by (\ref{eq:dif_erobab}) is not applicable and the probability distribution
of scattering assumes some different form. 
However, in most cases we need not worry about the effects of
this indeterminacy for small $\Delta H$ because 
usually we need only the integrated characteristics of the
probability function [e.g. in calculation of some scattering coefficients 
in Paper II which could be done using
$\tilde P(H,\Delta H)$, see below]. Integrals of 
$\tilde P(H,\Delta H)$
 are subject to strong cancellation 
effects near $\Delta H=0$ and this naturally removes the problem.
Similar situation was discussed by Goodman 
(1983, 1985)
in the context of stellar
gravitational encounters in globular clusters.

Using (\ref{eq:differ}) we can provide an alternative derivation
of $\langle \Delta \tilde h\rangle_{\tau,\omega}$.
Multiplying r.h.s. of (\ref{eq:differ}) by $\Delta H$ and integrating
from $\Delta H_-$ to $-\Delta H_{min}$ and 
from $\Delta H_{min}$ to 
$\Delta H_+$ one obtains that\footnote{Remember that we are using 
$\langle \Delta \tilde h\rangle$ and $\langle (\Delta \tilde h)^2\rangle$
instead of $\langle \Delta H\rangle$ and 
$\langle (\Delta H)^2\rangle$ for consistency with the notation of Paper II.}
\begin{eqnarray}
\langle \Delta \tilde h\rangle_{\tau,\omega}=-
\frac{4}{3\pi}\frac{
\tilde h\ln\left(|\Delta H_-|\Delta H_+/\Delta H_{min}^2\right)}
{\tilde i|H|\sqrt{\tilde e^2-H^2}
\left[\tilde e^2+\tilde i^2-(3/4)H^2\right]^{3/2}}.
\label{dh_new}
\end{eqnarray}
Using (\ref{eq:dh_em}) and (\ref{eq:dhmin}) we find that
\begin{eqnarray}
\ln\left(\frac{|\Delta H_-|\Delta H_+}{\Delta H_{min}^2}\right)=
2\ln\left(\tilde i\frac{v^2}{\Omega^2 R_H^2}\right)=2\ln\Lambda,
\label{delh_new}
\end{eqnarray}
which means that expression (\ref{dh_new}) coincides with equation (73)
of Paper II.

For some purposes it will still be necessary to use the 
distribution probability
(\ref{eq:dif_erobab}) itself rather than its integrals. For these occasions
we can remedy the divergence in (\ref{eq:dif_erobab}) by introducing some
cutoff distance $d\sim \Delta H_{min}$ into the denominator
of (\ref{eq:dif_erobab}). One can do this in a variety of ways, one of the 
simplest ones would be to take
\begin{eqnarray}
\tilde P(H,\tilde e,\tilde i,\Delta H)=\frac{1}{2\ln\Lambda}
\frac{\langle (\Delta \tilde h)^2\rangle_{\tau,\omega}
+\langle \Delta \tilde h\rangle_{\tau,\omega}\Delta H}
{(|\Delta H|+d)^3}.
\label{eq:dif_erobab_cutoff}
\end{eqnarray}
One can easily check that the normalization of the total probability to  
$1$ would require that
\begin{eqnarray}
d\approx \left[\frac{\langle (\Delta \tilde h)^2\rangle_{\tau,\omega}}
{2\ln\Lambda}\right]^{1/2}.
\label{eq:d_def}
\end{eqnarray}
We use this nondivergent form of (\ref{eq:dif_erobab_cutoff})
in \S \ref{subsubsect:pass_scattering}.

To take into account the distribution of planetesimal eccentricities and
inclinations we need to average the differential probability 
using Schwarzschild
velocity distribution. Expression (\ref{eq:dif_erobab}) provides 
a natural way of doing this and we find as a result that
\begin{eqnarray}
\tilde P(H,\tilde \sigma_e,\tilde \sigma_i,\Delta H)=
\frac{1}{2\ln\Lambda}
\frac{\langle (\Delta \tilde h)^2\rangle
+\langle \Delta \tilde h\rangle\Delta H}
{|\Delta H|^3},
\label{eq:dif_erobab_av}
\end{eqnarray}
where explicit analytic expressions for $\langle \Delta \tilde h\rangle$ and 
$\langle (\Delta \tilde h)^2\rangle$ are given by 
equations (86) and (87) of Paper II. In principle we should be
bearing in mind the restriction (\ref{eq:condit}) when doing this
last averaging. In practice this can only slightly 
affect the factor inside the 
logarithm  if we restrict the range of $\Delta H$ by approximate
condition
\begin{eqnarray}
\Delta \hat H_-<\Delta H <\Delta \hat H_+,~~~~~~
\Delta \hat H_\pm=-H\pm\sqrt{H^2+4 (\tilde \sigma_e^2+
\tilde \sigma_i^2-H^2)}.
\label{eq:condit_av}
\end{eqnarray}
Outside of this range simple form (\ref{eq:dif_erobab_av}) will give rather
poor approximation. The nondivergent form of (\ref{eq:dif_erobab_av})
can be obtained in the same fashion as (\ref{eq:dif_erobab_cutoff}) 
and (\ref{eq:d_def}).

%*************************************************************
%*************************************************************

\section{Intermediate velocity regime and distant encounters.}
\label{app:intermediate_velocity}

%*************************************************************
%*************************************************************

Equations (\ref{eq:surf_shear})-(\ref{eq:inc_shear}) and 
(\ref{eq:em_surf})-(\ref{eq:em_s_e})  describe the evolution 
of the disk properties
in the shear- and dispersion-dominated regimes correspondingly. Unfortunately
the assumptions we have made while deriving them preclude us from
using these equations  directly in the intermediate regime, when 
\begin{eqnarray}
\tilde \sigma_{e}^2+\tilde \sigma_{i}^2\sim 1.
\end{eqnarray}

Still we can try to provide an approximate description of evolution
in this velocity regime by smoothly interpolating between the
two extremes we believe we can describe well. 
Of course, one should not expect very good agreement
of the theory interpolated into the intermediate velocity regime with
the outcomes of orbit integrations. Qualitative concord would be 
enough for our purposes.

Far from the embryo where the scattering due to close encounters
is exponentially small (because of the scarcity of planetesimals 
with large eccentricities) distant encounters dominate
 the evolution of the planetesimal disk because their 
effect decays only as a power law in the embryo-planetesimal 
separation. We attempt to take them into account 
 by using the same interpolation 
technique: we assume that when $H/\tilde \sigma_e \ll 1$ close encounters
dominate and scattering is described by formulae of 
\S \ref{subsubsect:pass_scattering}. When $H/\tilde \sigma_e \gg 1$
distant encounters dominate and we use formulae of 
\S \ref{subsubsect:hs_scattering} to describe their effect on the 
disk properties because distant scattering is similar to the shear-dominated
scattering. 

To provide a smooth matching between the different velocity regimes 
and spatial scattering zones we
use  interpolating 
functions $\Theta(x)$ and $\Psi(x)$ with the following properties:
$\Theta(x), \Psi(x)\to 0$ as $x\to 0$ and 
$\Theta(x), \Psi(x) \to 1$ as $x\gg 1$.
Then we can describe evolution of some quantity $F$ ($N,\tilde \sigma_e^2,
\tilde \sigma_i^2$) in terms of the 
corresponding shear- and dispersion-dominated extremes as
\begin{eqnarray}
\frac{\partial F}{\partial t}=\Theta(\tilde \sigma_e^2+\tilde \sigma_i^2)
\frac{\partial F}{\partial t}\Bigg|_{disp}+\left[1-
\Theta(\tilde \sigma_e^2+\tilde \sigma_i^2)
\left(1-\Psi(H/\tilde \sigma_e)\right)\right]
\frac{\partial F}{\partial t}\Bigg|_{shear}.
\label{eq:interpolation}
\end{eqnarray}
The properties of $\Theta(x)$ and $\Psi(x)$ ensure that this formula
reproduces  correct limiting behaviors of disk evolution.

We have found that the results of orbit integrations performed 
in the intermediate velocity regime can be satisfactorily described using 
the following form of interpolating function $\Theta(x)$:
\begin{eqnarray}
\Theta(x)=\exp\left[-\frac{40}{x^{1/2}(5+x)^3}\right].
\label{eq:Theta_def}
\end{eqnarray}
For $\Psi(x)$ we use the following {\it ad hoc}
prescription:
\begin{eqnarray}
\Psi(x)=\exp\left[-(8/x)^4\right].
\label{eq:W_def}
\end{eqnarray}

\begin{figure}
\vspace{15cm}
\includegraphics{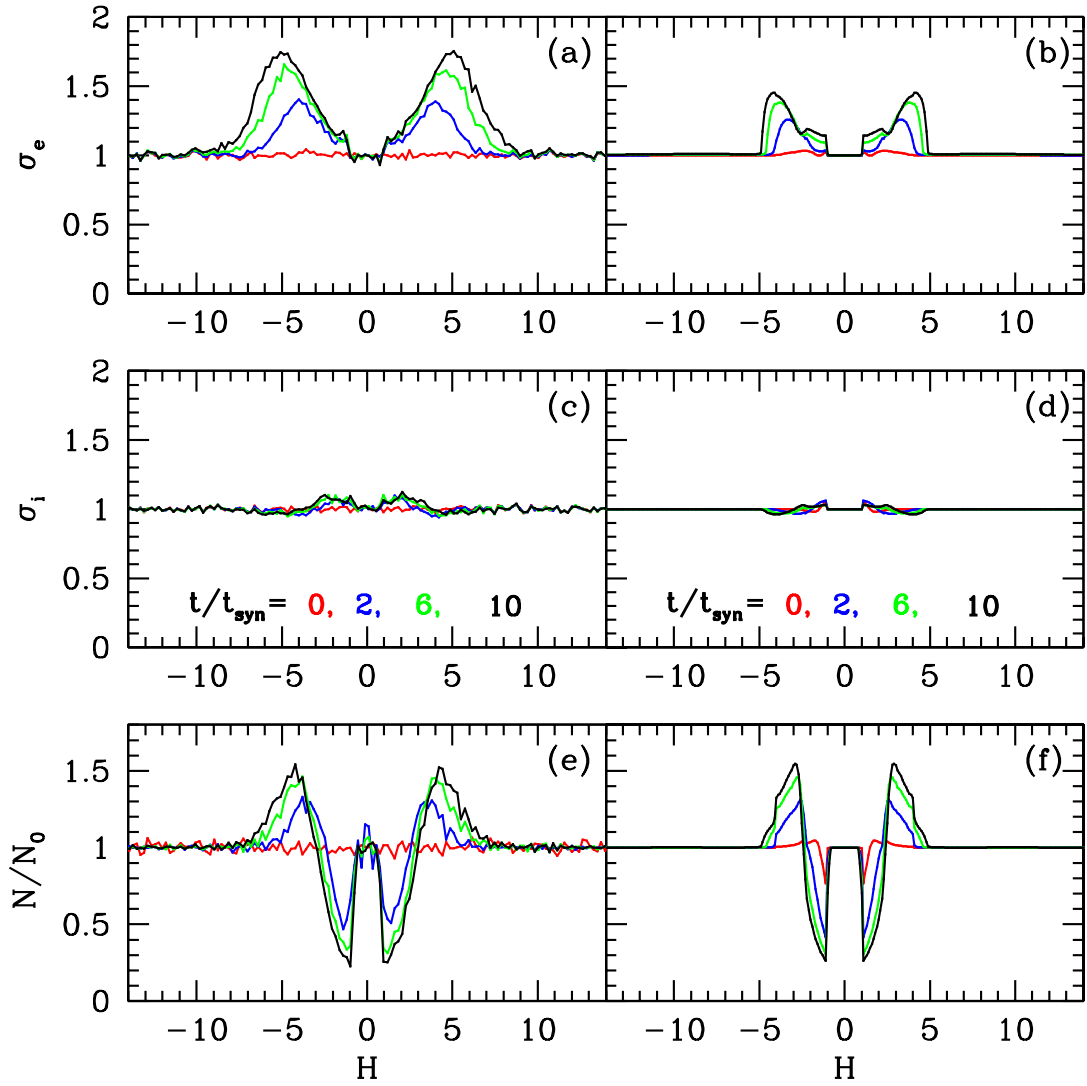}
\caption{The same as Figure \ref{fig:evolution_3_3} but for initial 
$\tilde \sigma_e=\tilde \sigma_i=1$. Dimensionless time is displayed
in panels {(c,d)}.}
\label{fig:evolution_0.3_0.3}
\end{figure}

Comparison of the disk evolution in the intermediate regime 
computed using equations
(\ref{eq:interpolation})-(\ref{eq:W_def}) with that obtained 
from direct orbit integrations is shown in Figure 
\ref{fig:evolution_0.3_0.3}. We display the results for a disk 
with initial $\tilde \sigma_e=\tilde \sigma_i=1$. 
One can see that the agreement between two methods 
is only qualitative indeed. 
The most probable reason for this is the 
use of the simple prescription (\ref{eq:surf_shear})-(\ref{eq:inc_shear})
with a discrete probability function (\ref{eq:delta_fun_cold})
to treat the shear-dominated regime. This also rules out the possibility 
of accurate treatment of the effects of distant encounters. 
In fact, as long as the random motion in the disk
in not exactly zero scattering always has some 
dispersion of outcomes and this gives rise to the 
 diffusive transport of
planetesimals in the disk. Inclusion of this effect would widen and lower 
the surface density profile as well as remove spiky features caused by
the discrete scattering function (\ref{eq:delta_fun_cold}). We do not 
pursue this subject here and leave its detailed exploration for the future. 
Still, we have verified using several different choices of $\tilde \sigma_e$
and $\tilde \sigma_i$ in the intermediate velocity regime that qualitative 
agreement is adequate: general features and trends appearing
in the numerical results are always reproduced reasonably 
well by our analytical prescription.

In Appendix \ref{app:accr_rate} we describe the calculation of the 
planetesimal accretion by the embryo in the shear- and dispersion-dominated 
regimes. We again use the interpolation approach for the description
of the accretion rate in the intermediate velocity regime:
\begin{eqnarray}
\dot M = \Pi(\tilde \sigma_e^2+\tilde \sigma_i^2)
\dot M_{disp}+\left[1-
\Pi(\tilde \sigma_e^2+\tilde \sigma_i^2)\right]
\dot M_{shear},
\label{eq:m_acc_interp}
\end{eqnarray}
where $\dot M_{disp}$ is given by (\ref{eq:accr}), 
$\dot M_{shear}$ by (\ref{eq:accr_shear}), and function 
$\Pi(x)$ has the same properties as $\Theta(x)$ and $\Psi(x)$.
We have chosen\footnote{This choice of $\Pi(x)$ certainly depends on 
the accuracy of formula (\ref{eq:accr_shear_DT}) and our 
specific value of $H_{coll}$, see Appendix \ref{app:accr_rate}.} 
\begin{eqnarray}
\Pi(x)=\exp\left[-\frac{1}{(2.8 x)^{1.45}}\right].
\label{eq:Pi_def}
\end{eqnarray}
Comparison of theoretical mass accretion rates obtained using 
(\ref{eq:m_acc_interp}) with those derived from the orbit 
integrations in the intermediate
dispersion regime is shown in Figure \ref{fig:acc_rate}c,d.

%*************************************************************
%*************************************************************

\section{Embryo accretion rate in an inhomogeneous planetesimal disk.}
\label{app:accr_rate}

%*************************************************************
%*************************************************************

The accretion rate of any massive body
immersed in a planetesimal disk depends sensitively on the dynamical 
state of the disk.  Detailed studies of various accretion regimes in
homogeneous disks were carried out  by a number of authors
(Greenzweig \& Lissauer 1990, 1992; Dones \& Tremaine 1993)
and we will use some of their results.

We will concentrate on one very important regime: when the maximum 
planetesimal impact parameter $l_{max}$ necessary for accretion to take place 
is much smaller than the disk vertical scaleheight. If this condition 
is not fulfilled the inclination of planetesimals must be 
unrealistically small (Dones \& Tremaine 1993),
\begin{eqnarray}
\tilde \sigma_i\le p,~~~~~p=\frac{R_e}{R_H}.
\label{eq:sigma_i_max}
\end{eqnarray}
The dimensionless parameter $p$ is the ratio of the 
embryo's radius $R_e$ to its
Hill radius $R_H$. It is independent of the mass of the accretor 
and is typically rather small in the Solar system ($< 10^{-2}$).
%Most likely, condition (\ref{eq:sigma_i_max}) does not hold true 
%during the late stages of the disk evolution 
%when embryos form. Then accretion has a well-defined
%$3$-dimensional character. The only other separation possible in this
%case is that due to the shear- or dispersion-dominated regimes.

In the dispersion-dominated regime,
when $\tilde\sigma_e^2+\tilde\sigma_i^2\gg 1$ [this is the $3$-dimensional
high-dispersion regime of Dones \& Tremaine (1993)],
 analytical treatment of the accretion 
rate is possible, because  the two-body approximation
is valid in this velocity regime (see Paper II). In this
approximation the accretion cross-section  is 
\begin{eqnarray}
S=\pi R_e^2\left[1+\frac{2G(M_e+m)}{R_e v^2}\right],
\label{eq:cross_sect}
\end{eqnarray}
where $M_e$ and $m$ are accretor and planetesimal masses, and $v$ is 
the approach velocity
of the planetesimal particles at infinity. The second term in this formula
takes into account gravitational focussing, which is very important 
in planetesimal disks.

We proceed in a manner analogous of that of Paper II
(see \S 4). We first fix $H,\tilde e,$ and
$\tilde i$ of planetesimals. Only planetesimals with $|H|<\tilde e$ 
can experience close encounters with the embryo and be accreted.
In addition, planetesimals on horseshoe orbits cannot be accreted either,
thus we put another restriction on $H$: $|H|>\tilde h_{hs}$, with
$\tilde h_{hs}$ given by (\ref{eq:pass_cond}).
Also, following the discussion in \S \ref{subsubsect:pass_scattering},
we need to distinguish the value of the semimajor axis difference 
 immediately before the encounter $H^\prime$ from its value at
infinity, $H$. The 
conservation of planetesimal flux means that the surface number densities
of planetesimals immediately before  the
encounter $N^\prime$ and far from the embryo $N$ are related via
\begin{eqnarray}
N^\prime(H^\prime) |H^\prime|dH^\prime=N(H) |H|dH.
\label{eq:relation}
\end{eqnarray}
If the dependence $H^\prime(H)$ is given by our simple approximation
(\ref{eq:h_hprime}) then $dH^\prime/dH=H/H^\prime$ and 
$N^\prime(H^\prime(H))=N(H)$;
of course this would not be true for any other relation between these
quantities. 

The total number of planetesimals with semimajor axes between $H$ and
$H+dH$, eccentricities and inclinations in the intervals
$[\tilde e,\tilde e+d\tilde e]$ and $[\tilde i,\tilde i+d\tilde i]$
correspondingly  passing the accretor in a unit of time is
\begin{eqnarray}
\delta N=\frac{3}{2}\Omega \frac{R_H^2}{a_e^2} 
N(H)\psi(\tilde e,\tilde i)|H|dH 
d\tilde e d\tilde i,
\label{eq:tot_num}
\end{eqnarray}
where  $\psi(\tilde e,\tilde i)$ is a distribution function of eccentricities 
and inclination for which we assume  a Gaussian form 
(\ref{eq:Railey}) following Paper II,
and $N$ is dimensionless.
The approach velocity of accreting planetesimals is
[equation (60) of Paper II]
\begin{eqnarray}
v^2=\Omega^2 R_H^2\left[\tilde e^2+\tilde i^2-(3/4)(H^\prime)^2\right].
\label{eq:approach_vel}
\end{eqnarray}
We have used $H^\prime$ in this formula because calculation of scattering 
in the two-body approximation uses values of the orbital
parameters immediately before the encounter.

The accretor will absorb all planetesimals which have impact parameters 
at infinity $l$ smaller than 
\begin{eqnarray}
l_{max}=R_e\sqrt{1+\frac{2G(M_e+m)}{R_e v^2}},
\label{eq:max_impact}
\end{eqnarray}
a condition from which (\ref{eq:cross_sect}) follows.
Then, using equation (\ref{eq:cumul}) we may write that
the fraction of $\delta N$ which can get accreted is
\begin{eqnarray}
f_a=P(H^\prime,l<l_{max})=\frac{2}{3\pi}\frac{l_{max}^2}{R_H^2}
\frac{v}{\Omega R_H}\frac{1}
{\tilde i|H^\prime|\sqrt{\tilde e^2-(H^\prime)^2}}.
\label{eq:fraction}
\end{eqnarray}
Then the mass accretion rate due to the population
of field particles with mass $m$ is 
\begin{eqnarray}
\dot M=m\int\limits_{-\infty}^\infty dH
\int\limits_{|H|}^\infty\int\limits_0^\infty d\tilde e d\tilde i 
\psi(\tilde e,\tilde i)
f_a\frac{d\delta N}{dH}\nonumber\\
=m\frac{\Omega R_e^2}{\pi a_e^2}\int\limits_{-\infty}^\infty dH
\int\limits_{|H|}^\infty\int\limits_0^\infty N(H)
\tilde v_0\frac{|H|}{|H^\prime|}
\frac{\psi(\tilde e,\tilde i)d\tilde e d\tilde i}
{\tilde i\sqrt{\tilde e^2-(H^\prime)^2}}
\left[1+\frac{2}{p \tilde v_0^2}\right],
\label{eq:gen_accr}
\end{eqnarray}
where we are using the following notation: 
$\tilde v(e,i)=v/(\Omega R_H)$, and $p$ is given by 
(\ref{eq:sigma_i_max}). 

We carry out the integration over $d\tilde e$ and $d\tilde i$ in a way 
similar to our calculation
of scattering coefficients in Appendix A of Paper II. As a 
result we obtain equations (\ref{eq:accr}) and (\ref{eq:Upm_def}).
Note that (\ref{eq:accr}) automatically takes into account the transition
between strong and weak gravitational focussing regimes which takes place at 
$\tilde\sigma_e,\tilde\sigma_i\sim p^{-1/2}\gg 1$.
We have numerically compared (\ref{eq:accr}) applied to a homogeneous
planetesimal disk (and $H^\prime=H$) with the corresponding analytical 
accretion rates of Greenzweig \& Lissauer (1992) and found them to agree.
The conversion between $H^\prime$ and $H$ turns out to be a significant
ingredient for the accuracy of the accretion rates; the use of simple
$H^\prime=H$ prescription instead of (\ref{eq:h_hprime})
leads to $\approx (10-20)\%$ bigger discrepancy
with our numerical results.

In the shear-dominated regime [``intermediate dispersion, strong gravity''
case of Dones \& Tremaine (1993)] the situation is different. 
The approach velocity
of planetesimals is always $\sim \Omega R_H$. Then the mass accretion rate
can depend on only the vertical (and not horizontal)
velocity dispersion since $\tilde\sigma_i$ determines
the disk thickness and local density of planetesimals. Using simple 
scaling arguments and  orbit integrations to fix constant coefficients 
Dones \& Tremaine (1993) have
 demonstrated that in the shear-dominated regime 
with the ratio of vertical to horizontal velocity dispersions 
$\tilde\sigma_i/\tilde\sigma_e=0.5$ the accretion rate 
in homogeneous disk is given by
\begin{eqnarray}
\dot M\simeq 5\frac{N m\Omega R_e R_H}{\tilde\sigma_i a_e^2}.
\label{eq:accr_shear_DT}
\end{eqnarray}

In our case this result is no longer applicable because planetesimal 
surface density is not the same in different parts of the disk.
In fact, when computing the 
accretion rate in inhomogeneous disk 
it is more meaningful to use {\it instantaneous} surface
number density of planetesimals on passing orbits which lead to 
collisions with the embryo.
We will assume that planetesimals on orbits near the horseshoe-passing boundary
end up colliding with the embryo in the cold regime
[see Petit \& H\'enon (1986) for more accurate locations of collision
bands in the shear-dominated regime]; this means that we
will be using the instantaneous surface density $N^{inst}
(H_{coll})$ where 
$H_{coll}\approx\pm 1.4$ [see (\ref{eq:my_cond})] 
as a measure of surface density 
in equation (\ref{eq:accr_shear_DT}). 
Using results of Paper II (Appendix B)
we can write that
\begin{eqnarray}
N^{inst}(H_{coll})=\frac{1}{\sqrt{2\pi}}\int
\limits^\infty_{-\infty}N(H)
\frac{dH}{\tilde\sigma_e(H)}
\exp\left[-\frac{(H-H_{coll})^2}{2\tilde\sigma_e^2(H)}\right].
\label{eq:N_inst}
\end{eqnarray}

All these considerations allow us to adopt
formula (\ref{eq:accr_shear}) for the accretion rate in the 
inhomogeneous disk 
with arbitrary ratio of 
vertical to horizontal planetesimal velocity dispersions.
Of course, in the 
homogeneous disk with $\tilde\sigma_i/\tilde\sigma_e=0.5$ 
it reproduces (\ref{eq:accr_shear_DT}).

In the intermediate velocity regime we smoothly interpolate between the 
accretion rates represented by formulae (\ref{eq:accr_shear}) 
and (\ref{eq:accr}), see Appendix \ref{app:intermediate_velocity}.
To compute the accretion rate in a disk with a distribution of planetesimal 
masses one simply needs to integrate our single-mass formulae over the
whole planetesimal mass spectrum.

\end{document}